\documentclass{article}

\PassOptionsToPackage{numbers, compress}{natbib}

\usepackage[final]{neurips}

\usepackage{latexsym}
\usepackage{tcolorbox}

\usepackage[T1]{fontenc}
\usepackage[utf8]{inputenc}
\usepackage{microtype}

\usepackage{inconsolata}
\usepackage{wrapfig,lipsum,booktabs}
\usepackage{pifont}
\usepackage{makecell} 
\usepackage{tabularx}
\usepackage[normalem]{ulem}
\useunder{\uline}{\ul}{}
\usepackage[export]{adjustbox}

\usepackage{hyperref}       
\usepackage{url}            
\usepackage{booktabs}       
\usepackage{amsfonts}       
\usepackage{nicefrac}       

\usepackage{graphicx}
\usepackage{subfigure}
\usepackage{float}
\usepackage{multirow}

\usepackage{listings}

\usepackage{amsmath}
\usepackage{amssymb}
\usepackage{mathtools}
\usepackage{amsthm}
\usepackage{balance}
\usepackage{flushend}

\usepackage{lipsum}
\usepackage{graphicx}
\usepackage{subcaption}
\usepackage{tikz}
\usetikzlibrary{shapes, arrows, positioning}

\usepackage{fancyhdr}

\hypersetup{
	colorlinks,
	citecolor=gray,
	linkcolor=red,
	urlcolor=blue}

\usepackage[capitalize,noabbrev]{cleveref}
\usepackage{tablefootnote}

\usepackage{color, colortbl,xcolor}
\usepackage{enumitem}
\usepackage{booktabs,arydshln}

\definecolor{bestRow}{HTML}{CFF5F1}
\usepackage{xspace}
\usepackage{xcolor}

\usepackage{natbib}
\usepackage{doi}

\newcommand{\TotalTrials}{4{,}590}
\newcommand{\TotalScenarios}{45}
\newcommand{\nbf}[1]{{\noindent \textbf{#1}}}

\def\paperTitle{\includegraphics[height=14pt]{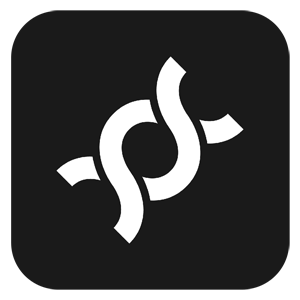} From Procedural Skills to Strategy Genes: Towards Experience-Driven Test-Time Evolution}

\title{\paperTitle}

\makeatletter
\def\adl@drawiv#1#2#3{%
	\hskip.5\tabcolsep
	\xleaders#3{#2.5\@tempdimb #1{1}#2.5\@tempdimb}%
	#2\z@ plus1fil minus1fil\relax
	\hskip.5\tabcolsep}
\newcommand{\cdashlinelr}[1]{%
	\noalign{\vskip\aboverulesep
		\global\let\@dashdrawstore\adl@draw
		\global\let\adl@draw\adl@drawiv}
	\cdashline{#1}
	\noalign{\global\let\adl@draw\@dashdrawstore
		\vskip\belowrulesep}}
\makeatother

\setlist[itemize]{align=parleft,left=0pt..0.5em}
\setlist[enumerate]{align=parleft,left=0pt..0.5em}

\setlist[itemize]{align=parleft,left=0pt..0.8em}

\usepackage{booktabs}
\newcolumntype{g}{>{\columncolor{airforceblue}}c}
\def\authorBlock{
    Junjie Wang$^\textnormal{2}$\footnotemark[1] \qquad
    Yiming Ren$^\textnormal{2}$\footnotemark[1] \qquad
    Haoyang Zhang$^\textnormal{1}$\footnotemark[1]
    \\\\\fontsize{10}{10}
    \selectfont{$^\textnormal{1}$Infinite Evolution Lab, EvoMap} \quad
    \selectfont{$^\textnormal{2}$Independent Researcher} \\
    {\tt\small 17@evomap.ai}
    
}

\author{\authorBlock}

\begin{document}

\maketitle

{
  \renewcommand{\thefootnote}%
  {\fnsymbol{footnote}}
  \footnotetext[1]{Equal contribution.}
}

\begin{abstract}
This beta technical report asks how reusable experience should be represented so that it can function as effective test-time control and as a substrate for iterative evolution. 
We study this question in \TotalTrials{} controlled trials across 45 scientific code-solving scenarios. We find that documentation-oriented Skill packages provide unstable control: their useful signal is sparse, and expanding a compact experience object into a fuller documentation package often fails to help and can degrade the overall average. 
We further show that representation itself is a first-order factor. 
A compact Gene representation yields the strongest overall average, remains competitive under substantial structural perturbations, and outperforms matched-budget Skill fragments, while reattaching documentation-oriented material usually weakens rather than improves it. 
Beyond one-shot control, we show that Gene is also a better carrier for iterative experience accumulation: attached failure history is more effective in Gene than in Skill or freeform text, editable structure matters beyond content alone, and failure information is most useful when distilled into compact warnings rather than naively appended. 
On CritPt, gene-evolved systems improve over their paired base models from 9.1\% to 18.57\% and from 17.7\% to 27.14\%. 
These results suggest that the core problem in experience reuse is not how to supply more experience, but how to encode experience as a compact, control-oriented, evolution-ready object.
\end{abstract}

\section{Introduction}

\begin{figure}[h]
\centering
\includegraphics[width=0.8\textwidth]{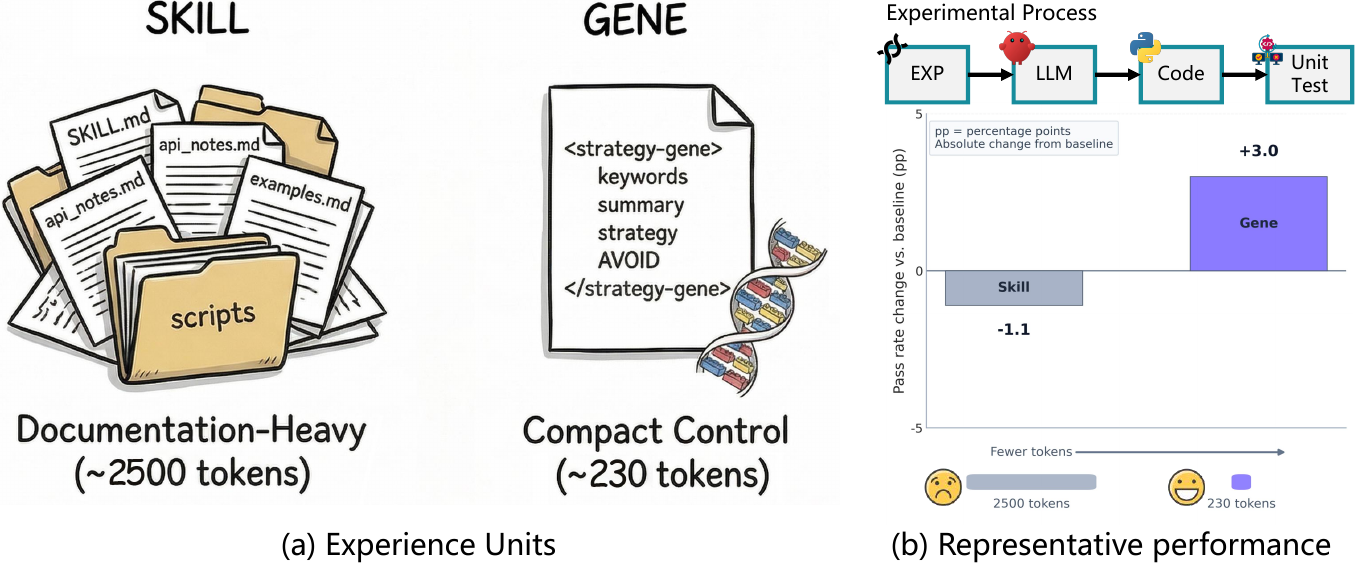}
\caption{
Experience units for test-time control and their representative performance.
(a) Skill as a documentation-heavy experience package ($\sim2{,}500$ tokens) versus Gene as a compact control-oriented representation ($\sim230$ tokens).
(b) Representative results relative to the no-guidance baseline in the evaluation pipeline: Gene yields a $+3.0$pp gain, whereas Skill incurs a $-1.1$pp drop.
}
\label{fig:teaser}
\end{figure}

Large language model (LLM) agents are increasingly designed to improve through the accumulation and reuse of prior experience. 
Unlike one-shot prompt engineering, a growing body of work equips agents with mechanisms such as reflection, reusable experience repositories, and structured skill artifacts to preserve useful information from past interactions and reuse it in subsequent problem solving~\cite{auto-agent-survey,LLM-based-agent}.
Such experience may take the form of verbal reflection, external memory, executable skills, or experience libraries~\cite{Voyager,Reflexion,ExpeL}. 
Although these approaches differ in implementation, most of them treat experience primarily as a content object that can be stored, retrieved, and replayed. 
In contrast, this paper further asks whether such experience can in fact function as a stable and effective control signal at test time.

This assumption is further reinforced by recent skill-style approaches. 
Rather than preserving experience as free-form textual feedback or isolated case fragments, these methods tend to organize experience into procedural units with explicit interfaces, applicability conditions, and execution structure. 
ProcMEM formulates experience as executable skills with explicit execution structure~\cite{ProcMEM}, while MemSkill further treats memory operations themselves as learnable and evolvable skills~\cite{MemSkill}. 
Memento-Skills~\cite{Memento-Skills} and CoWork-X~\cite{CoWork-X} push this trend further by organizing experience as persistent, structured, and compositional skill libraries to support reliable and budget-aware improvement.
Taken together, these studies reinforce a stronger assumption about reusable experience: more complete and structured experience representations should be more beneficial for subsequent task solving.

Our point of departure is to test this assumption directly. 
As shown in~\cref{fig:teaser}(b), a representative comparison already suggests that more experience content does not necessarily translate into better test-time control. 
A compact \texttt{Gene} representation, at roughly $230$ tokens, improves performance by $+3.0$pp over the baseline, whereas a documentation-heavy \texttt{Skill} package, at roughly $2{,}500$ tokens, reduces performance by $-1.1$pp. 
This contrast shifts attention from experience quantity to experience representation: \emph{what form of reusable experience can serve as an effective and evolvable test-time control signal for LLM-based task solving?}

Building on this observation, this paper reexamines the experience object itself. 
We treat a \textbf{procedural skill} as a documentation-oriented experience object that is useful for human reading, instruction, and review, but not necessarily well suited to serve as a directly usable control signal under limited inference budget. 
In contrast, the \textbf{strategy gene} proposed in this work is a compact, structured, and behavior-oriented representation of experience. 
Its objective is not documentary completeness, but higher signal density, clearer applicability boundaries, and stronger control relevance within a constrained token budget. 
Accordingly, a strategy gene is not merely a compressed version of a procedural skill, but a different abstraction of reusable experience. 
To support stable accumulation and iterative revision of such objects, we introduce the Gene Evolution Protocol (GEP) as a protocol layer for canonicalizing genes into structured, evolvable experience representations.

To study this object-level shift, we conduct \TotalTrials{} controlled experiments on complex scientific tasks, spanning $45$ scenarios, and design three analytical probes. 
The \textbf{Skill Probe} examines why procedural skills often fail to provide stable and effective test-time control. 
The \textbf{Gene Probe} investigates why strategy genes can support experience-based control with lower budget and greater robustness. 
The \textbf{Evolution Probe} asks which properties make strategy genes suitable as the substrate for continual accumulation and experience-driven evolution. 
Taken together, these probes are not intended to compare a set of prompt tricks in isolation, but to systematically test a representational shift from \emph{documentation-oriented experience objects} to \emph{control-oriented, evolution-ready experience objects}.

Our main findings are as follows. 
(1) \emph{Procedural skills} are often poorly aligned with the requirements of test-time control: only a narrow subset of high-density procedural content contributes meaningfully, while much of the surrounding documentation imposes additional burden. 
(2) Representation is itself a first-order factor. Even when the underlying experience content is held broadly constant, substantial differences still arise from how that experience is packaged, structured, and exposed to the model. 
(3) \emph{Strategy genes} provide a more token-efficient and control-oriented experience representation than full skill-style documents, suggesting that effective reusable experience is not equivalent to longer instruction. 
(4) Experience accumulation is most effective when it remains selective and structured: compact warnings, editable schema, and stable carriers are more useful than additive attachment of diffuse history.

The contributions of this paper are threefold:
\begin{enumerate}[leftmargin=1.5em,itemsep=0pt,topsep=2pt]
    \item We recast reusable experience from a problem of storing and invoking content into one of representing test-time control signals.
    \item Through controlled probes, we identify several object-level factors governing effective experience reuse, including information overload in documentation-oriented skills, representational packaging effects, structural robustness, bounded reuse, and selective experience accumulation.
    \item We introduce strategy genes, organized under the Gene Evolution Protocol (GEP), as a protocolized control representation for reusable experience, and show in scientific code-solving settings that they provide a stronger interface than documentation-oriented skill objects for test-time control and iterative evolution.
\end{enumerate}

\section{Related Work}
\label{sec:related_work}

\subsection{Experience Reuse in Agents and Test-Time Improvement}

A growing body of work improves LLM agents by enabling them to accumulate and reuse experience across tasks. 
One line of research focuses on test-time refinement through critique, repair, and reflection, where models iteratively improve outputs using self-generated, execution-grounded, or tool-assisted feedback rather than parameter updates~\cite{Self-Refine,CRITIC,Self-Debug}. 
Another line externalizes experience into persistent memory or reusable repositories, such as episodic verbal memory, natural-language lessons, and executable skill libraries that can be recalled in later tasks~\cite{Reflexion,ExpeL,Voyager}. 
Although these approaches differ in mechanism, they largely share the same assumption: useful experience can be preserved as a reusable content object and later replayed to improve subsequent problem solving. 
Our work is motivated by a prior question left open by this literature: once experience is preserved, what representation should it take when it re-enters the model at test time so that it functions as stable and effective control rather than merely additional context?

\subsection{Experience Representation, Control, and Evolution}

Recent work has begun to make reusable experience more structured, compact, and maintainable. 
Procedural-memory and skill-centric approaches increasingly encode experience as explicit skill units with executable structure, learnable operations, or persistent external skill files~\cite{Memp,ProcMEM,MemSkill,Memento-Skills}. 
Related systems further organize such units into structured and compositional skill libraries under reliability and budget constraints, or extract reusable expertise patterns from trajectories to support continual refinement~\cite{CoWork-X,AutoRefine}. 
A closely related line studies how externalized experience should be maintained and evolved during deployment, emphasizing streaming updates, joint optimization of extraction and management, and even meta-evolution of memory architectures~\cite{Evo-Memory,TAME,UMEM,MemEvolve}. 
Taken together, these works increasingly recognize that reusable experience should be compact, structured, and amenable to continual refinement. 
However, their primary focus remains on how experience can be accumulated, maintained, or evolved once externalized. 
By contrast, our focus is on a prior representational question: what kind of experience object is most suitable for acting as a test-time control signal in the first place? 
This shift leads us from documentation-oriented experience objects toward compact, failure-aware, and protocolized control representations.

\section{Experience Representations for Test-Time Control}
\label{sec:representation}

We study reusable experience as an \emph{experience representation}: an externalized object derived from prior problem solving and reintroduced at test time to influence model behavior. 
This perspective shifts the focus from whether experience can be stored and recalled to what representational form allows it to function as effective test-time control. 
It also motivates a distinction between documentation-oriented representations, optimized for human use, and control-oriented representations, optimized for model-facing inference under limited budget.

\subsection{Reusable Experience Representations}
\label{sec:rer}

Let $\mathcal{H}=\{\tau_i\}_{i=1}^{n}$ denote a set of prior problem-solving trajectories. 
We define a \emph{reusable experience representation} as an externalized object
\begin{equation}
r=\phi(\mathcal{H}), \qquad r \in \mathcal{R},
\end{equation}
where $\phi$ abstracts prior experience into a reusable representation space $\mathcal{R}$.

Given a new task input $x$, a fixed model with parameters $\theta$ produces
\begin{equation}
y \sim p_{\theta}(\cdot \mid x, r).
\end{equation}
A representation is control-relevant only if it induces a measurable shift in test-time behavior:
\begin{equation}
p_{\theta}(y \mid x, r) \neq p_{\theta}(y \mid x, \varnothing)
\end{equation}
for a non-negligible subset of tasks under the target distribution.

This definition excludes three nearby cases. 
Reusable experience representations are not parameter updates, since $\theta$ remains fixed; not one-off prompt tricks, since they are abstracted from prior experience and intended for reuse; and not mere conversational recall, since their role is to provide task-relevant control rather than preserve dialogue continuity. 
Accordingly, the central question is not only how experience is stored or retrieved, but what representational form allows prior experience to act as effective test-time control.

\subsection{Procedural Skills}
\label{sec:skills}

A \emph{procedural skill} is a documentation-oriented experience representation that organizes prior problem-solving knowledge into a human-readable and reusable package~\cite{SkillsBench,AgentSkills}.
It is designed primarily for interpretation, instruction, review, and archival, rather than for direct behavioral control under constrained inference-time budget.

Formally, we model a procedural skill as a document package
\begin{equation}
s = \{d_{\text{main}}, d_{\text{aux}}\},
\end{equation}
where
\begin{equation}
d_{\text{main}} = \{\texttt{overview}, \texttt{workflow}, \texttt{pitfalls}, \texttt{error\_handling}\}
\end{equation}
contains the core procedural description, and
\begin{equation}
d_{\text{aux}} = \{\texttt{api\_notes}, \texttt{examples}, \texttt{scripts}\}
\end{equation}
contains supplementary reference and execution-support material.
This structure captures the documentation-heavy form of skill-style experience used in our experiments, where the instantiated \texttt{Skill} representation is approximately $2{,}500$ tokens long.

Under this view, procedural skills are valuable because they externalize experience in a stable and communicable form, making prior solutions easier to document, teach, maintain, and transfer across tasks.
However, these advantages do not imply suitability for inference-time control.
Procedural skills are optimized for readability and knowledge transfer, whereas test-time control requires compact, behaviorally targeted guidance that remains effective under limited token budget and constrained attention.
A representation that is useful for human understanding may therefore be inefficient or even misaligned when placed in model context.
Whether documentation-oriented skill packages can serve as effective control representations is thus an empirical question rather than a premise~\cite{AGENTS.md}.

\subsection{Strategy Genes and the GEP Protocol}
\label{sec:genes}

We define a \emph{strategy gene} as a control-oriented experience representation distilled from prior problem-solving experience. 
Unlike procedural skills, which package experience for reading, explanation, and review, a strategy gene is designed to deliver control-relevant experience under constrained inference budget. 
Its objective is not documentary completeness, but compactness, structural clarity, behavioral targeting, and failure awareness. 
Accordingly, a strategy gene is not a shortened skill, but a different abstraction of reusable experience.

Formally, let $s$ denote a skill-style experience package and let $\mathcal{H}$ denote a set of prior problem-solving trajectories. 
A strategy gene is obtained by a distillation map
\begin{equation}
g=\psi(s)\quad \text{or} \quad g=\psi(\mathcal{H}), \qquad g \in \mathcal{G},
\end{equation}
where $\psi$ extracts a compact control-oriented representation from a richer source experience object. 
At the representation level, we model a gene as
\begin{equation}
g=(m,u,\pi,\alpha,c,v),
\end{equation}
where $m$ denotes task-matching signals, $u$ a compact summary, $\pi$ a small set of strategic steps, $\alpha$ a set of failure-aware \texttt{AVOID} cues, $c$ optional execution constraints, and $v$ optional validation hooks. 
In our operational setup, this corresponds to a compact template containing keywords, summary, strategy, and \texttt{AVOID} fields, with optional constraint and validation metadata.

This structure differs from procedural skills in both organization and purpose. 
Skills organize experience according to documentation logic; genes organize experience according to control logic. 
Skills prioritize readability and completeness; genes prioritize signal density, applicability boundaries, and failure salience. 
The intended use is therefore different as well: a procedural skill is primarily a documentation artifact, whereas a strategy gene is a model-facing control representation.

This distinction also explains why genes require protocolization. 
Without a protocol layer, a gene remains a free-form prompt fragment whose boundaries are unstable and whose internal fields are difficult to compare, manipulate, or accumulate systematically. 
We therefore introduce the \emph{Gene Evolution Protocol} (GEP) as a representation interface that canonicalizes genes into structured objects:
\begin{equation}
\tilde{g}=\Gamma(g), \qquad \tilde{g}\in\widetilde{\mathcal{G}},
\end{equation}
where $\Gamma$ maps a raw gene instance into a canonical protocol-compliant form. 

In the GEP view, genes are the atomic capability units, while higher-level objects such as capsules and events capture validated execution paths and immutable evolution logs, respectively. 
Our focus in this paper is the gene layer\footnote{A reference implementation for transforming procedural skills into GEP-compliant representations (including genes) is available at \url{https://github.com/EvoMap/skill2gep}.}, because it is the minimal reusable unit that can directly function as test-time control, while still supporting composition and evolution under a shared protocol. 
A fuller protocol-level formalization of GEP, including genes, capsules, events, and the protocol loop, is provided in~\cref{app:gep}.

Canonicalization through GEP serves three purposes. 
First, it provides stable boundaries and internal structure, making genes explicit objects rather than ad hoc text snippets. 
Second, it makes genes comparable and operable: once protocolized, genes can be matched, replaced, revised, and composed at the object level. 
Third, it provides the minimal interface required for accumulation and evolution, so that experience units can be inherited, corrected, validated, and iteratively refined over time. 
In this sense, GEP is not a formatting detail, but the protocol layer that lifts a control-oriented representation into an evolution-ready experience object. 
If procedural skills are documentation artifacts, strategy genes under the GEP protocol are evolution-ready control representations.

\subsubsection{Experimental Settings}

In this paper, \emph{test-time control} refers to the ability of an externalized experience representation to alter model behavior at inference time without modifying model parameters.
Formally, given a task input $x$ and a reusable experience representation $r$, a fixed model with parameters $\theta$ generates
\begin{equation}
y \sim p_{\theta}(\cdot \mid x, r; \gamma),
\end{equation}
where $\gamma$ denotes the inference configuration.
In our experiments, $\gamma$ is held fixed across conditions, using low-temperature decoding ($T=0.05$), a maximum output length of $16{,}384$ tokens, and a shared prompt injection interface.
Operationally, the task description is supplied separately from the auxiliary control representation, allowing the latter to function as an explicit test-time control signal.

We focus on scientific tasks whose outputs are executable Python programs evaluated in a sandbox.
For trial $i$, let $p_i$ denote the number of checkpoints passed by the generated program and $n_i$ the total number of checkpoints for that scenario.
We define the trial-level pass rate as
\begin{equation}
r_i = \frac{p_i}{n_i},
\end{equation}
and report condition-level pass rate as the average trial-level score:
\begin{equation}
\mathrm{PassRate} = \frac{1}{N}\sum_{i=1}^{N} r_i.
\end{equation}
A complete pass is additionally recorded when all checkpoints succeed, i.e., $p_i=n_i$, but this is treated as a secondary descriptive statistic rather than the primary metric.
Each trial is executed under a timeout of $120$ seconds.

\begin{figure}[t]
\centering
\includegraphics[width=\textwidth]{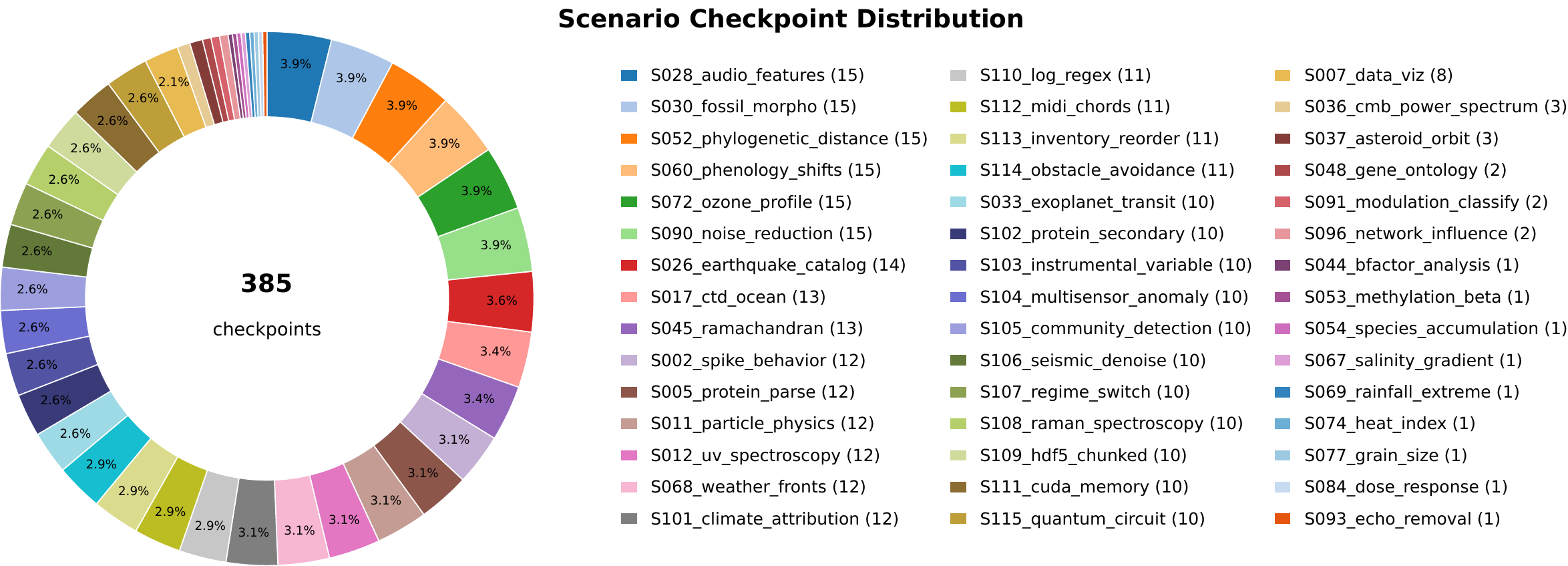}
\caption{Scenario-level checkpoint distribution of the benchmark used in this paper. 
The benchmark contains scenarios with substantially different checkpoint granularities, ranging from low-granularity end-to-end tasks to high-granularity multi-step tasks. 
}
\label{fig:scenario_checkpoint_distribution}
\end{figure}

All experiments use two fixed Gemini models, Gemini 3.1 Pro Preview (Pro) and Gemini 3.1 Flash Lite Preview (Flash), and comprise \TotalTrials{} retained trials in total.
``Avg.'' denotes the arithmetic mean over Pro and Flash.
$\Delta$ is computed against the no-guidance baseline within this experiment, in percentage points.
As shown in~\cref{fig:scenario_checkpoint_distribution}, benchmark scenarios vary substantially in checkpoint granularity, which is why we evaluate conditions using checkpoint-based pass rate rather than binary trial success alone.
Representative tasks include protein parsing, UV-Vis spectroscopy peak detection, exoplanet transit analysis, earthquake catalog processing, climate attribution, community detection, and inventory reordering.
We provide the detailed settings in~\cref{app:exp_setup}.

\subsubsection{Running Example}

We use \texttt{S012\_uv\_spectroscopy} as a running example.
The task requires generating a Python program that reads UV-Vis spectra from CSV files, detects peaks, computes peak attributes such as wavelength, height, FWHM, and area, identifies the dominant peak for each sample, and writes structured outputs.
Representative checkpoints cover command-line interface correctness, peak detection, wavelength and height extraction, FWHM and area computation, dominant-peak identification, and output structure.

This scenario is particularly suitable for illustrating test-time control because several recurring errors arise not from missing high-level task understanding, but from incorrect operationalization of library behavior.
Two representative failure modes are to treat \texttt{min\_distance} as a wavelength value rather than converting it into sample-index units before calling \texttt{scipy.signal.find\_peaks}, and to use the raw output of \texttt{peak\_widths} without converting it back to wavelength units before reporting FWHM.

The corresponding \texttt{Skill} representation packages this experience in a documentation-oriented form, combining task overview, workflow description, and auxiliary reference material.
The corresponding \texttt{Gene} representation packages the same underlying experience in a compact control-oriented form that foregrounds a small number of high-risk decision points.
A gene instance is shown below.

\begin{quote}
\small
\texttt{<strategy-gene>} \\
\texttt{Domain keywords: uv-vis, peak detection, FWHM, unit conversion} \\
\texttt{Summary: Detect peaks and compute wavelength-domain peak properties correctly} \\
\texttt{Strategy:} \\
\texttt{1. Detect peaks with prominence-based criteria} \\
\texttt{2. Convert \texttt{min\_distance} into sample-index units before peak detection} \\
\texttt{3. AVOID: Report FWHM only after converting \texttt{peak\_widths} outputs back to wavelength units} \\
\texttt{</strategy-gene>}
\end{quote}

Under checkpoint-based evaluation, these differences affect not only whether the final program fully passes, but also which sub-tasks succeed.
For example, incorrect handling of \texttt{min\_distance} may reduce peak-count accuracy while leaving file parsing and output formatting intact, whereas incorrect conversion of \texttt{peak\_widths} specifically degrades FWHM-related checkpoints.
This example illustrates the central comparison of the paper: the same underlying experience can produce different test-time control effects depending on whether it is packaged for documentation or for behavioral guidance.

\section{Results and Analysis}

This section presents the empirical analyses of Gene-Bench from three perspectives: the effect of documentation-style Skill guidance, the representational properties of Gene as a control interface, and the role of Gene as a carrier for accumulating experiential information. 

\subsection{Skill Probe: Documentation-Oriented Skills Are Misaligned with Test-Time Control}

This probe asks whether documentation-oriented \texttt{Skill} prompts function as effective test-time control, and if not, where their usable control value actually resides.
This probe includes 1,440 retained trials in total.

\subsubsection{Overall Comparison}

As shown in~\cref{fig:teaser}(b) and~\cref{tab:skill_probe_overall}, we begin with the overall comparison between no guidance, \texttt{Gene}, and the full \texttt{Skill} package, all derived from the same per-scenario experience source.

\begin{table}[h]
\centering
\small
\caption{Overall comparison of no guidance, \texttt{Gene}, and \texttt{Skill}. Avg. denotes the mean of Pro and Flash; $\Delta$ is computed against no guidance.}
\label{tab:skill_probe_overall}
\begin{tabular}{lcccc}
\toprule
Condition & Pro & Flash & Avg. & $\Delta$ \\
\midrule
\texttt{Skill} & 50.7\% & 49.0\% & 49.9\% & -1.1 \\
No guidance & 60.1\% & 41.8\% & 51.0\% & 0.0 \\
\cellcolor{bestRow}\texttt{Gene} & \cellcolor{bestRow}59.9\% & \cellcolor{bestRow}48.2\% & \cellcolor{bestRow}54.0\% & \cellcolor{bestRow}+3.0 \\
\bottomrule
\end{tabular}
\end{table}

\nbf{\texttt{Gene} is the strongest condition on the overall average.}
\texttt{Gene} reaches $54.0\%$ ($+3.0$pp), exceeding no guidance at $51.0\%$ and \texttt{Skill} at $49.9\%$ ($-1.1$pp).
This shows that, when the same underlying experience is recast as a compact control-oriented object, it transfers more effectively than the full documentation package.
Although \texttt{Skill} raises Flash from $41.8\%$ to $49.0\%$, it simultaneously lowers Pro from $60.1\%$ to $50.7\%$.
Its benefit is therefore not robust at the benchmark level, and the full documentation package does not function as a reliable test-time control interface.

\nbf{\texttt{Gene} preserves performance more cleanly than \texttt{Skill}.}
Relative to no guidance, \texttt{Gene} remains nearly unchanged on Pro ($59.9\%$ vs.\ $60.1\%$) while still improving Flash to $48.2\%$.
Compared with \texttt{Skill}, it gives up little on Flash but avoids the large degradation on Pro.
This is the expected profile of a control-facing experience object rather than a human-facing document.

\begin{figure*}[t]
\centering
\includegraphics[width=\textwidth]{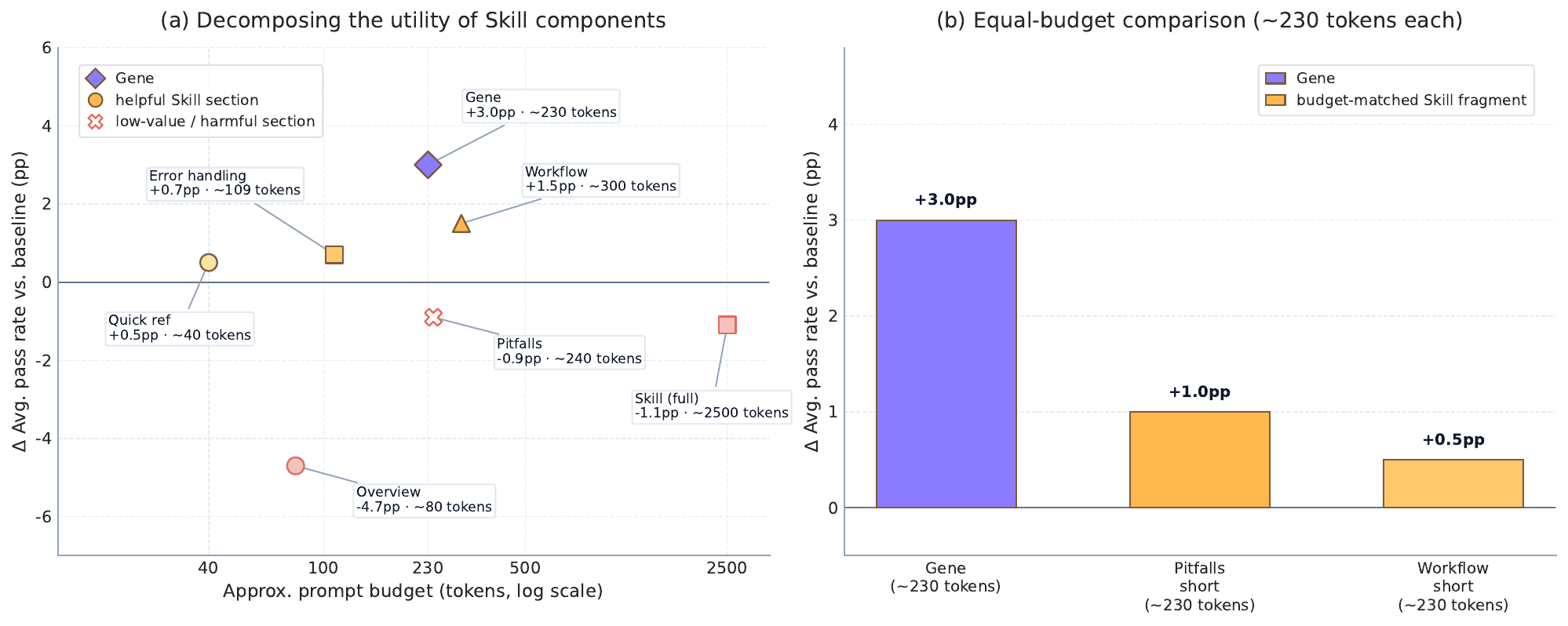}
\caption{\textbf{\texttt{Skill}'s control value is sparse, whereas \texttt{Gene} remains stronger even under matched budget.}
(a) Decomposing \texttt{Skill} shows that only a narrow procedural slice is clearly useful, while several sections are neutral or harmful.
(b) Under an approximately matched budget, shortened \texttt{Skill} fragments improve substantially, yet still remain below \texttt{Gene}.}
\label{fig:skill_decompose_equal_budget}
\end{figure*}

\subsubsection{Localizing the Usable Control Signal within \texttt{Skill}}

We next decompose \texttt{Skill} into individual sections and compare budget-matched \texttt{Skill} fragments against \texttt{Gene}. 
Additional numerical details are provided in~\cref{app:skill_sections} and~\cref{app:skill_budget_matched}.

\cref{fig:skill_decompose_equal_budget} summarizes the main pattern. Panel~(a) shows that the utility of \texttt{Skill} is highly uneven across sections. 
\texttt{Skill-Workflow} is the only clearly useful component, whereas \texttt{Skill-Overview} is strongly harmful, and the full \texttt{Skill} package remains below the no-guidance baseline. 
This indicates that the usable control signal within \texttt{Skill} is not broadly distributed across the document, but concentrated in a narrow procedural slice.

\nbf{The effective control content in \texttt{Skill} is sparse and predominantly procedural.}
Only \texttt{Skill-Workflow} yields a clear positive gain, whereas descriptive sections such as \texttt{Skill-Overview} substantially reduce performance. 
The full \texttt{Skill} package also underperforms the baseline. 
These results indicate that most of the document does not operate as actionable test-time control, and that the usable signal is concentrated in workflow-style guidance.

\nbf{\texttt{Gene} is stronger than full \texttt{Skill} and stronger than most individual \texttt{Skill} sections.}
Within the same decomposition setting, \texttt{Gene} achieves the largest gain overall, exceeding the full \texttt{Skill} package as well as \texttt{Skill-QuickRef}, \texttt{Skill-ErrorHandling}, and \texttt{Skill-Pitfalls}. 
Only \texttt{Skill-Workflow} remains competitive among the isolated sections. This suggests that \texttt{Gene} is not merely a shortened version of a single document fragment, but a more effective packaging of the usable control signal.

Panel~(b) then examines whether this advantage can be explained purely by prompt budget. 
When \texttt{Skill} is aggressively trimmed to approximately the same budget as \texttt{Gene}, the shortened fragments improve substantially over the full \texttt{Skill} package. 
This confirms that part of \texttt{Skill}'s weakness is indeed attributable to packaging overhead.

\nbf{Reducing \texttt{Skill} to the same budget narrows the gap, yet does not reverse the conclusion.}
Under matched budget, shortened \texttt{Skill} fragments become markedly more competitive. 
However, \texttt{Gene} remains the strongest condition overall. 
The difference is therefore not reducible to brevity alone. Rather, what matters is how experience is organized as a control-facing object.

\nbf{The main limitation of \texttt{Skill} lies not in missing knowledge, but in how that knowledge is presented.}
\texttt{Skill} does contain useful information, most clearly in procedural guidance. 
However, that value is sparse and readily diluted by surrounding documentation-oriented material. 
\texttt{Gene} remains stronger because it packages the relevant experience more directly for inference-time control.

\subsection{Gene Probe: Strategy Genes Are a Better Representation of Reusable Experience}

We next examine whether \texttt{Gene} is merely a shorter prompt, or a genuinely better representation of reusable experience. This probe comprises 1,890 retained trials.

\subsubsection{Construction of \texttt{Gene}}

We begin with a controlled construction from no guidance to the full \texttt{Gene}, progressively adding keywords, summary, and strategy. 
\cref{tab:gene_progressive_construction} shows that the benefit of \texttt{Gene} does not follow a simple token-budget trend. 

\begin{table}[h]
\centering
\small
\caption{Progressive construction of \texttt{Gene}. Avg. is the mean of Pro and Flash; $\Delta$ is relative to no guidance.}
\label{tab:gene_progressive_construction}
\begin{tabular}{lcccc}
\toprule
Condition & Pro & Flash & Avg. & $\Delta$ \\
\midrule
No guidance & 60.1\% & 41.8\% & 51.0\% & 0.0 \\
\texttt{Gene} (keywords + summary) & 51.3\% & 50.6\% & 51.0\% & +0.0 \\
\texttt{Gene} (keywords only) & 57.9\% & 49.1\% & 53.5\% & +2.5 \\
\cellcolor{bestRow}\texttt{Gene} (keywords + summary + strategy) & \cellcolor{bestRow}59.9\% & \cellcolor{bestRow}48.2\% & \cellcolor{bestRow}54.0\% & \cellcolor{bestRow}+3.0 \\
\bottomrule
\end{tabular}
\end{table}

\nbf{The benefit of \texttt{Gene} does not follow a simple token-budget trend; it emerges only when experience is organized into strategy.}
The keywords-only variant reaches $53.5\%$, the keywords+summary variant falls back to the baseline level at $51.0\%$ , and the full \texttt{Gene} reaches the highest overall average at $54.0\%$. The gain is therefore not explained by progressively adding more tokens. Rather, the key improvement appears only when the representation reaches the strategy layer, where prior experience is organized into a more explicit control interface.

\nbf{\texttt{Gene} is not simply the shortest effective prompt.}
The keywords-only variant already performs well, yet the full \texttt{Gene} remains the strongest condition overall. This rules out a trivial interpretation in which any sufficiently short prompt would work. What distinguishes \texttt{Gene} is not merely compactness, but the way compact experience is structured into actionable guidance.

\subsubsection{Robustness to Content and Structural Perturbations}
\label{sec:robustness_content_structural}

\begin{figure}[h]
\centering
\includegraphics[width=\textwidth]{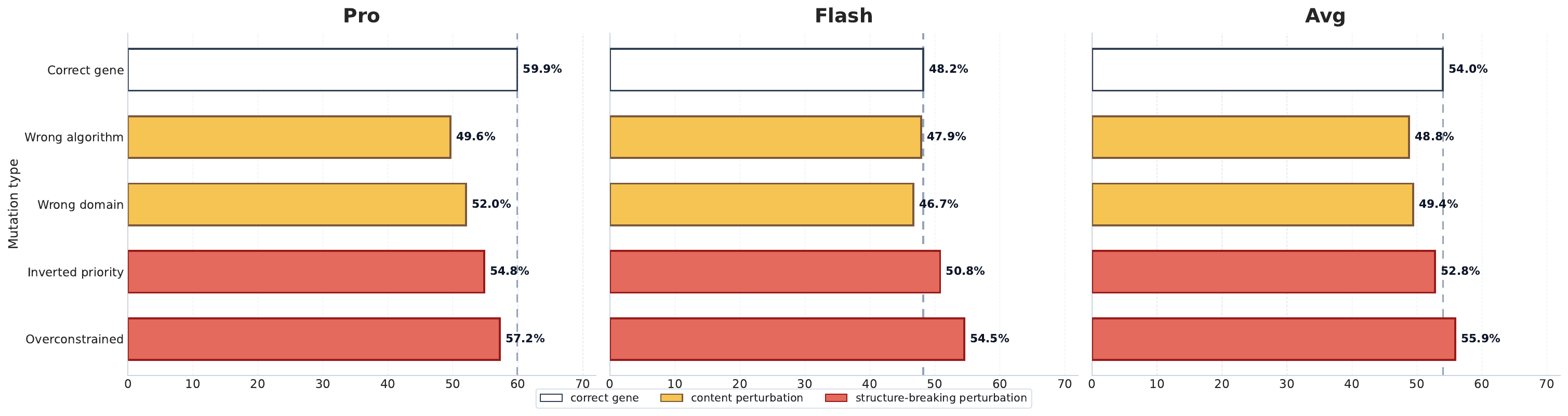}
\caption{\textbf{\texttt{Gene} is substantially more sensitive to content corruption than to structural distortion.}
Wrong algorithm and wrong domain reduce performance on both Pro and Flash, whereas inverted priority remains competitive and overconstrained guidance even improves over clean \texttt{Gene} in this setting. This suggests that \texttt{Gene}'s effect is not tied to one fixed surface form, but depends more strongly on whether the encoded experience remains task-appropriate.}
\label{fig:gene-mutation}
\end{figure}

We next test whether the effect of \texttt{Gene} is tied to one specific wording, or whether it remains stable under substantial perturbations. 
\cref{fig:gene-mutation} summarizes the main pattern. Two observations are immediate. First, content corruption is consistently harmful: replacing the correct strategy with a wrong algorithm or a wrong domain reduces Avg.\ from $54.0\%$ to $48.8\%$ and $49.4\%$, respectively. Second, structural perturbations are much less damaging. Inverted priority remains relatively close to clean \texttt{Gene} at $52.8\%$, and the overconstrained variant further rises to $55.9\%$.

\nbf{\texttt{Gene} is robust to substantial structural deformation.}
Neither reversing the priority order nor making the guidance overly restrictive collapses performance.
On the contrary, both structure-breaking variants remain competitive with, and in one case exceed, clean \texttt{Gene}.
This argues against a fragile surface-form explanation in which \texttt{Gene} works only when expressed in one canonical template.

\nbf{Semantic corruption is more damaging than structural distortion.}
When the perturbation changes the underlying experience rather than its presentation, performance drops much more sharply.
Wrong algorithm reduces Avg.\ to $48.8\%$, and wrong domain to $49.4\%$, both well below clean \texttt{Gene} at $54.0\%$.
This indicates that what matters most is preserving task-relevant control content, not preserving a fixed textual arrangement.

\nbf{The representation is robust, but not content-agnostic.}
These results do not support the view that \texttt{Gene} works merely because it contains a few generally useful tokens.
However, they also do not imply that any mutation is harmless.
The pattern is more specific: \texttt{Gene} tolerates substantial variation in structure, yet remains sensitive to whether the encoded experience is semantically appropriate for the task.

\subsubsection{Adding Documentation Back to \texttt{Gene}}

We test whether \texttt{Gene} can be further improved by reattaching selected documentation-oriented material.
\cref{tab:gene_selective_complementarity} shows that \texttt{Gene} alone remains the strongest condition overall. 

\begin{table}[h]
\centering
\small
\caption{Adding documentation-oriented material back to \texttt{Gene}. Avg. is the mean of Pro and Flash; $\Delta$ is relative to no guidance.}
\label{tab:gene_selective_complementarity}
\begin{tabular}{lcccc}
\toprule
Condition & Pro & Flash & Avg. & $\Delta$ \\
\midrule
\texttt{Skill} & 50.7\% & 49.0\% & 49.9\% & -1.1 \\
No guidance & 60.1\% & 41.8\% & 51.0\% & 0.0 \\
\texttt{Gene} + \texttt{API notes} & 51.8\% & 51.2\% & 51.5\% & +0.5 \\
\texttt{Gene} + \texttt{examples} & 57.8\% & 46.1\% & 52.0\% & +1.0 \\
\cellcolor{bestRow}\texttt{Gene} & \cellcolor{bestRow}59.9\% & \cellcolor{bestRow}48.2\% & \cellcolor{bestRow}54.0\% & \cellcolor{bestRow}+3.0 \\
\bottomrule
\end{tabular}
\end{table}

\nbf{Adding documentation back to \texttt{Gene} usually dilutes rather than complements it.}
When API notes are added to \texttt{Gene}, performance drops from $54.0\%$ to $51.5\%$; when examples are added, it drops to $52.0\%$. Neither augmented condition surpasses \texttt{Gene} alone. This indicates that the relation between \texttt{Gene} and \texttt{Skill} is not one of simple complementarity. Once a compact control-oriented object is expanded with documentation-oriented material, the additional text tends to weaken rather than strengthen control.

\nbf{The advantage of \texttt{Gene} is representational rather than additive.}
If \texttt{Gene}'s strength came primarily from being incomplete but improvable, then adding relevant auxiliary material should have produced further gains. However, the opposite pattern is observed. The strongest condition remains \texttt{Gene} alone, which suggests that its advantage comes from how experience is represented and exposed to the model, rather than from the amount of attached documentation.

\subsubsection{Composition of Multiple \texttt{Gene}s}

We examine whether \texttt{Gene} remains effective when multiple instances are composed within the same prompt.
\cref{tab:gene_multi_combination} shows that the best condition is not multi-\texttt{Gene} composition, but a single targeted \texttt{Gene}.

\begin{table}[h]
\centering
\small
\caption{Composition of multiple \texttt{Gene}s. Avg. is the mean of Pro and Flash; $\Delta$ is relative to no guidance.}
\label{tab:gene_multi_combination}
\begin{tabular}{lcccc}
\toprule
Condition & Pro & Flash & Avg. & $\Delta$ \\
\midrule
Two complementary \texttt{Gene}s & 45.5\% & 44.3\% & 44.9\% & -6.1 \\
Three complementary \texttt{Gene}s & 54.5\% & 46.2\% & 50.4\% & -0.6 \\
No guidance & 60.1\% & 41.8\% & 51.0\% & 0.0 \\
Two conflicting \texttt{Gene}s & 57.1\% & 49.4\% & 53.2\% & +2.2 \\
\cellcolor{bestRow}Single \texttt{Gene} & \cellcolor{bestRow}59.9\% & \cellcolor{bestRow}48.2\% & \cellcolor{bestRow}54.0\% & \cellcolor{bestRow}+3.0 \\
\bottomrule
\end{tabular}
\end{table}

\nbf{The benefit of \texttt{Gene} is bounded under composition, especially in specialized scientific settings.}
In our benchmark, the strongest condition remains a single targeted \texttt{Gene}, whereas combining multiple \texttt{Gene}s either weakens the gain or turns it into a loss.
A plausible explanation is that these tasks require relatively precise and domain-specific control.
Under such conditions, adding multiple \texttt{Gene}s—even when they are nominally complementary—may blur the control focus, introduce competing constraints, or weaken professional specificity.
The result therefore should not be read as evidence that \texttt{Gene} is inherently non-compositional, but rather that naive composition does not reliably work in high-difficulty scientific scenarios.

\nbf{Complementary composition is more harmful than conflicting composition.}
Two complementary \texttt{Gene}s yield the worst overall result at $44.9\%$, substantially below both no guidance and all other multi-\texttt{Gene} settings. By contrast, two conflicting \texttt{Gene}s remain relatively competitive at $53.2\%$. This suggests that the main failure mode is not contradiction alone. Rather, multiple partially relevant control objects may compete for attention and jointly blur the intended control signal, even when they are nominally compatible.

\nbf{Reusability therefore has a scope boundary.}
The three-\texttt{Gene} complementary condition partially recovers from the collapse of the two-\texttt{Gene} complementary setting, but still remains below the single-\texttt{Gene} condition. Taken together, these results suggest that \texttt{Gene} is reusable, but not in an unconstrained bag-of-skills sense. Its effectiveness depends on selecting a sufficiently targeted control object, rather than accumulating multiple related objects within the same inference context.

\subsection{Evolution Probe: Genes Provide a Better Substrate for Experience-Driven Evolution}

We next examine whether \texttt{Gene} provides a more suitable carrier for attaching and reorganizing accumulated experience over time. This probe comprises 1,260 retained trials.

\subsubsection{Carrier Format for Attached Failure History}

We first compare different carriers for attaching failure history, including \texttt{Gene}, \texttt{Skill}, and freeform text. 
\cref{tab:evolution_failure_carriers} shows that carrier format matters substantially. 

\begin{table}[h]
\centering
\small
\caption{Carrier format for attached failure history. Avg. is the mean of Pro and Flash; $\Delta$ is relative to no guidance.}
\label{tab:evolution_failure_carriers}
\begin{tabular}{lcccc}
\toprule
Condition & Pro & Flash & Avg. & $\Delta$ \\
\midrule
\texttt{Skill} + \texttt{failure} & 53.8\% & 41.8\% & 47.8\% & -3.2 \\
\texttt{Freeform} + \texttt{failure} & 55.7\% & 43.5\% & 49.6\% & -1.4 \\
No guidance & 60.1\% & 41.8\% & 51.0\% & 0.0 \\
\texttt{Gene} + \texttt{failure} & 55.3\% & 48.6\% & 52.0\% & +1.0 \\
\cellcolor{bestRow}\texttt{Gene} & \cellcolor{bestRow}59.9\% & \cellcolor{bestRow}48.2\% & \cellcolor{bestRow}54.0\% & \cellcolor{bestRow}+3.0 \\
\bottomrule
\end{tabular}
\end{table}

\nbf{\texttt{Gene} is a better carrier for attached experience than \texttt{Skill} or freeform text.}
When the same type of failure history is appended to different carriers, the \texttt{Gene}-based condition remains clearly stronger than both \texttt{Skill}+\texttt{failure} and \texttt{Freeform}+\texttt{failure}. This suggests that accumulated experience is not carrier-neutral: a structured control-facing object preserves attached information more effectively than either documentation-oriented or unstructured text.

\nbf{However, attaching failure history naively can still dilute \texttt{Gene}.}
Although \texttt{Gene}+\texttt{failure} is the best failure-attached condition, it remains below \texttt{Gene} alone by $2.0$pp on Avg. This indicates that the issue is not only which carrier is used, but also how new experience is incorporated. Appending failure text without further compression or reorganization can weaken an otherwise effective control object.

\subsubsection{Editable Structure versus Flattened Prose}

We next compare \texttt{Gene} in its native structured form against the same content flattened into prose. 
\cref{tab:evolution_editable_vs_static} isolates the effect of representation form while keeping content approximately fixed. 

\begin{table}[h]
\centering
\small
\caption{Editable structure versus flattened prose. Avg. is the mean of Pro and Flash; $\Delta$ is relative to no guidance.}
\label{tab:evolution_editable_vs_static}
\begin{tabular}{lcccc}
\toprule
Condition & Pro & Flash & Avg. & $\Delta$ \\
\midrule
\texttt{Gene}, flattened prose & 56.8\% & 44.1\% & 50.5\% & -0.5 \\
No guidance & 60.1\% & 41.8\% & 51.0\% & 0.0 \\
\cellcolor{bestRow}\texttt{Gene}, structured & \cellcolor{bestRow}59.9\% & \cellcolor{bestRow}48.2\% & \cellcolor{bestRow}54.0\% & \cellcolor{bestRow}+3.0 \\
\bottomrule
\end{tabular}
\end{table}

\nbf{Editable structure matters beyond content alone.}
Because the two conditions contain the same underlying experience, the gap cannot be explained by content availability or token count. 
The structured GEP form itself contributes to performance, indicating that editability and explicit schema are not merely cosmetic properties, but part of what makes \texttt{Gene} an effective carrier.

\nbf{Flattening \texttt{Gene} into prose removes most of its advantage.}
Once the structured object is converted into flowing text, its gain largely disappears. 
This shows that the benefit of \texttt{Gene} is not reducible to what it says alone; it also depends on how the experience is exposed to the model.

\subsubsection{Encoding Failure History within \texttt{Gene}}

We finally examine how failure information should be encoded when both prior failure and correct strategy are available.
\cref{tab:evolution_failure_encoding} shows that the best condition is neither of the mixed variants, but \texttt{failure warnings only}, which reaches $54.4\%$ ($+4.6$pp). 

\begin{table}[h]
\centering
\small
\caption{Encoding failure history within \texttt{Gene}. Avg. is the mean of Pro and Flash; $\Delta$ is relative to no guidance.}
\label{tab:evolution_failure_encoding}
\begin{tabular}{lcccc}
\toprule
Condition & Pro & Flash & Avg. & $\Delta$ \\
\midrule
No guidance & 57.9\% & 41.8\% & 49.8\% & 0.0 \\
\texttt{Failure first} & 56.3\% & 44.7\% & 50.5\% & +0.7 \\
\texttt{Strategy first} & 58.4\% & 45.2\% & 51.8\% & +2.0 \\
\texttt{Strategy only} & 56.9\% & 47.7\% & 52.3\% & +2.5 \\
\cellcolor{bestRow}\texttt{Failure warnings only} & \cellcolor{bestRow}56.8\% & \cellcolor{bestRow}52.0\% & \cellcolor{bestRow}54.4\% & \cellcolor{bestRow}+4.6 \\
\bottomrule
\end{tabular}
\end{table}

\nbf{Failure history is most effective when distilled into standalone warnings.}
The strongest condition in this analysis is \texttt{failure warnings only}, not either mixed ordering. This suggests that the most useful form of accumulated failure experience is not a long composite object, but a compact and explicit warning signal.

\nbf{Combining failure and strategy naively weakens both.}
Both \texttt{failure first} and \texttt{strategy first} underperform not only \texttt{failure warnings only}, but also \texttt{strategy only}. The problem is therefore not just ordering. Rather, once both signals are packed together, the resulting object becomes less effective than either distilled component on its own.

\nbf{Experience accumulation should be selective rather than additive.}
Taken together with the previous two analyses, this result suggests that evolution does not help by simply attaching more history to an existing object. The effective path is more selective: compressing failure into focused warnings, preserving structured editability, and avoiding additive growth that blurs control.

\subsection{Gene as a Carrier for Test-Time Evolution}
\label{sec:gene_tte}

\begin{figure}[h]
\centering
\includegraphics[width=\textwidth]{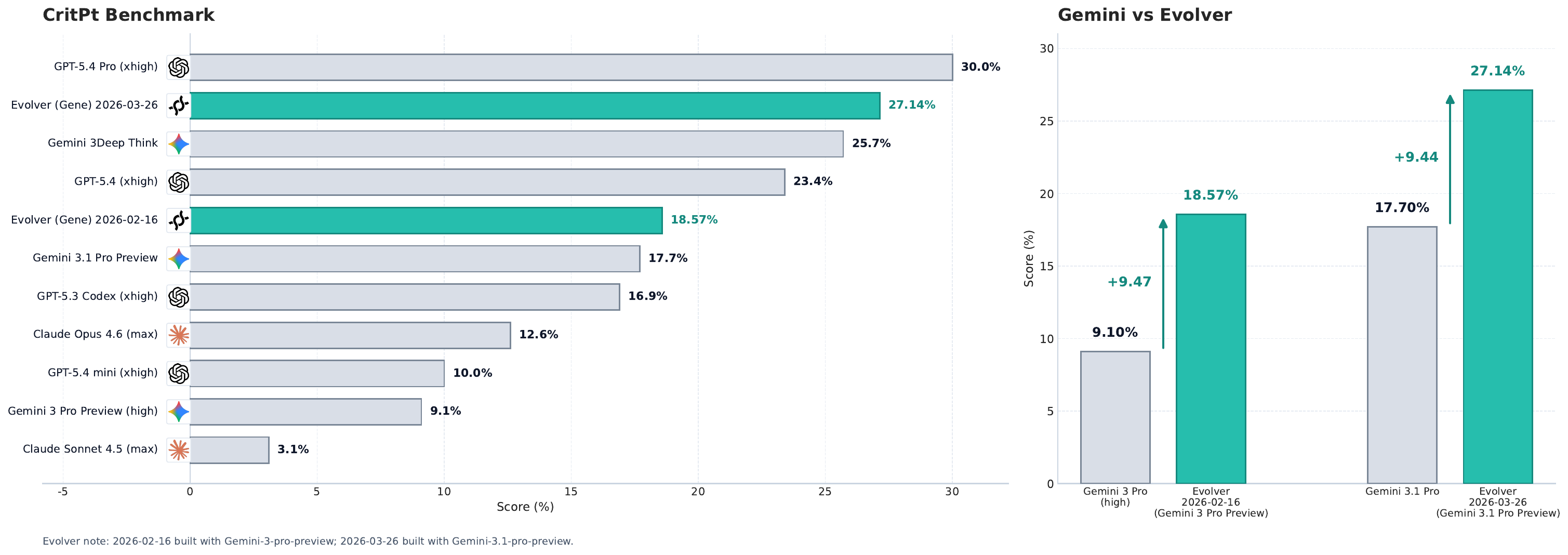}
\caption{
Accuracy (\%) on the CritPt benchmark.
Two gene-evolved systems, \texttt{Evolver (Gene) 2026-02-16} and \texttt{Evolver (Gene) 2026-03-26}, substantially outperform their paired base models, Gemini 3 Pro Preview and Gemini 3.1 Pro Preview, by $+9.47$ and $+9.44$ percentage points, respectively.
The stronger evolved version reaches $27.14\%$ accuracy, illustrating that gene-based test-time evolution can yield substantial gains beyond the underlying base model.
}
\label{fig:critpt_benchmark}
\end{figure}

We next ask a stronger question than one-shot test-time control: can \emph{Gene} also serve as a carrier for \emph{test-time evolution}, enabling iterative improvement across runs without changing the underlying base model?
To study this, we construct an evolutionary agent using OpenClaw\footnote{\url{https://github.com/openclaw/openclaw}} as the host runtime and Evolver\footnote{\url{https://github.com/EvoMap/evolver}} as the evolution engine.
At a high level, Evolver organizes self-evolution around structured Genes, accumulates causal experience through memory, and solidifies successful updates through validation-governed evolution events.
As shown in~\cref{fig:critpt_benchmark}, both evolved agents substantially outperform their paired base models on CritPt benchmark~\cite{CritPt}: \texttt{Evolver (Gene) 2026-02-16} reaches $18.57\%$ versus $9.1\%$ for Gemini 3 Pro Preview, while \texttt{Evolver (Gene) 2026-03-26} reaches $27.14\%$ versus $17.7\%$ for Gemini 3.1 Pro Preview.
Details are deferred to~\cref{app:critpt_evolution_details}.

\nbf{Version A: \texttt{Evolver (Gene) 2026-02-16}.}
We view the earlier version as \emph{memory-grounded} evolution: its gains mainly come from consolidating prior failures, execution traces, and corrective experience into reusable control units.
A representative case is \texttt{gene\_gep\_repair\_from\_errors}, which is triggered by explicit failure signals such as \texttt{error}, \texttt{exception}, \texttt{failed}, and \texttt{unstable}, and follows a tightly constrained repair loop: structured diagnosis, blast-radius estimation, smallest reversible patch, validation, and solidification back into the Gene/Capsule store.
The key point is that the gene does not merely remind the agent to ``be careful.'' It packages a reusable repair procedure.
Once distilled, the same pattern can be invoked again in later runs instead of being rediscovered from scratch.
This is the first sense in which gene iteration is useful: it converts past failures into reusable future-time control.

\nbf{Version B: \texttt{Evolver (Gene) 2026-03-26}.}
The later version reflects a broader, \emph{exploration-augmented} regime.
Across $70$ tasks, the system uses $210$ gene slots and $36$ unique gene IDs, with \texttt{arxiv}-derived genes accounting for $148$ selections, compared with $44$ from \texttt{run\_experience} and $18$ from \texttt{builtin\_topic\_prior}. 
The most revealing case is \texttt{gene\_topic\_hamiltonian\_inverse\_design}, the most frequent high-value run-experience gene in the full run, selected $25$ times.
Its content is already strongly procedural: enumerate constraints such as commutation, normalization, and operator ordering; construct coefficients under those constraints; decompose many-body-chain problems into local terms while preserving index consistency; and prioritize numerical stability and reproducibility, ideally with symbolic and numerical checking.
This shows what gene iteration preserves at a later stage: not just memories of failure, but compact task-facing solution procedures.
Its practical value is visible in repeated high-frequency combinations, especially the recurring trio \texttt{gene\_topic\_hamiltonian\_inverse\_design + gene\_topic\_many\_body\_spin\_chain + gene\_topic\_seed\_many\_body\_spin\_chain}, which appears across 12 tasks.
This is the second sense in which gene iteration is useful: the system is no longer only repairing itself, but also accumulating directly reusable solution patterns and deploying them repeatedly at scale.

\nbf{Discussion.}
Taken together, the two versions show that gene-based evolution is not merely becoming deeper, but broader.
\texttt{Evolver (Gene) 2026-02-16} mainly consolidates failures into stable repair and validation loops, whereas \texttt{Evolver (Gene) 2026-03-26} further converts successful task experience into reusable domain-facing procedures.
This distinction is central to our overall argument.
If a reusable experience unit is to support test-time evolution, it must remain compact enough for inference-time injection, structured enough for targeted update, and practical enough to preserve procedures that continue to work in later problems.
The CritPt results provide concrete evidence for this view: Gene is not only a control-oriented representation for one-shot inference, but also a persistent optimization interface through which reusable experience can accumulate, improve, and become increasingly useful through iteration.

\subsection{Overall Findings}

The retained analyses support four main findings.

\begin{itemize}[leftmargin=1.2em, itemsep=0.25em, topsep=0.2em, parsep=0pt, partopsep=0pt]

    \item \nbf{Documentation-oriented skills are misaligned with test-time control.}
    The useful control signal inside \texttt{Skill} is sparse and concentrated in a narrow procedural slice. Expanding a compact experience object into a fuller documentation package does not reliably help and can degrade the overall average.

    \item \nbf{Strategy genes are better understood as control-oriented experience objects rather than merely shorter prompts.}
    The benefit of \texttt{Gene} does not follow a simple token-budget trend, but emerges when experience is organized into strategy. Matched-budget \texttt{Skill} fragments narrow the gap but do not reverse the conclusion, and reattaching documentation-oriented material usually weakens rather than improves \texttt{Gene}.

    \item \nbf{\texttt{Gene} provides a better substrate for experience accumulation, but accumulation should be selective rather than additive.}
    Attached failure history is more effective in \texttt{Gene} than in \texttt{Skill} or freeform text, and editable structure matters beyond content alone. At the same time, naively appending additional failure text can dilute an otherwise effective control object.

    \item \nbf{Failure experience is most useful when distilled into compact warnings, and gene-based evolution can accumulate increasingly useful strategy objects over time.}
    Within \texttt{Gene}, compact failure warnings outperform mixed strategy--failure bundles, indicating that accumulation works best through selective compression rather than additive growth. Beyond one-shot control, gene-evolved systems also achieve substantial gains over their paired base models, showing that genes can support iterative improvement as a persistent optimization interface.

\end{itemize}

\section{Conclusion}
\label{sec:conclusion}

This paper studies a representational question in experience reuse: not how to provide more prior experience at test time, but how to encode reusable experience so that it functions as effective control under inference constraints. Across controlled experiments, we show that documentation-oriented \texttt{Skill} objects are poorly aligned with test-time control: their useful signal is sparse and often diluted by surrounding documentation-oriented material.
Against this background, we introduce \texttt{Gene} as a compact control-oriented experience representation. The retained analyses show that its advantage is not reducible to shorter prompting alone. Rather, \texttt{Gene} provides a more effective interface for organizing strategy, preserving task-relevant control, and supporting bounded reuse.
Beyond one-shot control, we further show that \texttt{Gene} is a better substrate for iterative experience accumulation. Attached failure history is more effective in \texttt{Gene} than in \texttt{Skill} or freeform text, and failure information is most useful when distilled into compact warnings rather than naively appended. Across two evolutionary runs on CritPt, gene-evolved systems substantially outperform their paired base models, suggesting that compact and revisable experience objects can support iterative improvement over time.
More broadly, our results suggest that experience reuse in LLM-based systems should be studied at the level of representation design rather than prompt packaging alone. We hope this work motivates further research on control-oriented experience objects, bounded reuse, and evolution-ready representations for test-time adaptation.

{
\bibliographystyle{unsrtnat}
\bibliography{references}
}

\clearpage
\appendix
\section*{Appendix}

\section{Gene Evolution Protocol (GEP)}
\label{app:gep}

\begin{table}[h]
\centering
\caption{GEP object schema. Gene follows the operational schema used in our experiments; capsule and event are paper-side formalizations of higher-level protocol objects.}
\label{tab:gep_schema}
\begin{adjustbox}{max width=\textwidth}
\begin{tabular}{p{1.3cm} p{2.5cm} p{7.0cm} p{3.8cm}}
\toprule
\textbf{Object} & \textbf{Field} & \textbf{Meaning} & \textbf{Role} \\
\midrule

\multirow{9}{*}{Gene}
& \texttt{type} & Object type, e.g., \texttt{Gene} & Canonical typing \\
& \texttt{schema\_version} & Protocol schema version & Compatibility / governance \\
& \texttt{id} & Unique gene identifier & Stable reference \\
& \texttt{signals\_match} & Keywords or trigger cues & Task matching \\
& \texttt{summary} & Compact one-sentence description & High-level intent \\
& \texttt{strategy} & Short ordered strategy list, including \texttt{AVOID} items & Core control logic \\
& \texttt{constraints} & Optional execution constraints & Bounded behavior \\
& \texttt{validation} & Optional executable checks or validation hooks & Testability \\
& \texttt{asset\_id} & Canonical hash / asset fingerprint & Provenance / deduplication \\
\midrule

\multirow{6}{*}{Capsule}
& $q$ & Task signature or problem context & Task grounding \\
& $G_{\kappa}$ & Genes instantiated in the episode & Strategy provenance \\
& $T_{\kappa}$ & Execution trace / action path & Realized procedure \\
& $o_{\kappa}$ & Observed outcome & Result record \\
& $V_{\kappa}$ & Validation / audit record & Verified execution \\
& $\ell_{\kappa}$ & Lineage pointer & Evolution linkage \\
\midrule

\multirow{8}{*}{Event}
& $t$ & Event type (e.g., repair, optimize, validation\_pass) & Transition label \\
& $\rho$ & Run or episode identifier & Episode traceability \\
& $a_{\mathrm{src}}$ & Source asset & Mutation origin \\
& $a_{\mathrm{dst}}$ & Target asset & Mutation result \\
& $\sigma$ & Triggering signal & Why evolution starts \\
& $\iota$ & Intended objective & Mutation intent \\
& $\Delta$ & Proposed diff / change & Transformation record \\
& $\nu, \tau$ & Validation result and timestamp & Auditability / ordering \\
\bottomrule
\end{tabular}
\end{adjustbox}
\end{table}

We provide a formalization of the Gene Evolution Protocol (GEP) used in this work. 
As shown in~\cref{tab:gep_schema}, consistent with the public GEP description, we treat the protocol as an object layer for reusable experience rather than as a mere prompt format or logging convention. 
At a high level, GEP organizes experience into three object types---\emph{genes}, \emph{capsules}, and \emph{events}---and updates them through a trial--validation--solidification loop. 
In the Gene-Bench setting used in this paper, the gene layer is the primary object of study, while capsules and events define the audit and evolution interfaces around it.

\subsection{Protocol Scope and Design Philosophy}
\label{app:gep_scope}

GEP is designed to support three requirements that free-form prompt fragments do not satisfy reliably. 
First, reusable experience should have stable boundaries and a canonical internal structure, so that it can be analyzed and reused as an explicit object rather than as unstructured text. 
Second, experience objects should be comparable and operable: they should admit matching, replacement, revision, composition, and validation. 
Third, experience should remain evolvable over time, meaning that successful adaptations can be accumulated, inherited, corrected, and audited rather than merely replayed as one-off prompt patches.

Under this view, GEP is not introduced to make prompting more elaborate. 
It is introduced to make reusable experience \emph{object-like}: explicit, structured, and protocolized enough to support lifecycle management at test time and across repeated task episodes.

\subsection{Object Hierarchy}
\label{app:gep_objects}

GEP defines three object layers:
\begin{equation}
\mathcal{O} = \mathcal{G} \cup \mathcal{C} \cup \mathcal{E},
\end{equation}
where $\mathcal{G}$ denotes the set of genes, $\mathcal{C}$ the set of capsules, and $\mathcal{E}$ the set of events.

\nbf{Genes.}
A \emph{gene} is the atomic capability unit. 
It is the smallest reusable experience representation that can directly function as test-time control. 
In this paper, genes are compact strategy templates distilled from richer experience sources such as procedural skills or prior trajectories.

\nbf{Capsules.}
A \emph{capsule} is a validated task-level execution path with audit trail. 
Whereas a gene encodes an abstract reusable strategy, a capsule records how one or more genes are instantiated and validated in a concrete task-solving episode. 
Capsules therefore operate at a higher granularity than genes: they capture successful compositions, execution context, and validation outcomes.

\nbf{Events.}
An \emph{event} is an immutable evolution log. 
Events record protocol-relevant state transitions, such as mutation, repair, validation failure, validation success, or solidification. 
Their role is not to serve as direct control input, but to preserve provenance, lineage, and auditability of experience evolution.

This hierarchy induces the following functional distinction:
\begin{itemize}
    \item \textbf{gene}: control unit
    \item \textbf{capsule}: validated execution unit
    \item \textbf{event}: evolution record
\end{itemize}

\subsection{Gene Objects}
\label{app:gep_gene}

A gene is a compact, control-oriented representation distilled from prior experience:
\begin{equation}
g=\psi(z), \qquad z \in \{s,\mathcal{H},\mathcal{C}\}, \quad g\in\mathcal{G},
\end{equation}
where $z$ may be a procedural skill $s$, a set of trajectories $\mathcal{H}$, or a previously validated capsule $\mathcal{C}$.

For the purposes of this paper, we formalize a gene as
\begin{equation}
g=(m,u,\pi,\alpha,c,v,\eta),
\end{equation}
where
\begin{itemize}
    \item $m$ denotes task-matching signals (e.g., keywords or trigger cues),
    \item $u$ denotes a compact summary of the intended behavior,
    \item $\pi$ denotes a small set of strategic steps,
    \item $\alpha$ denotes failure-aware \texttt{AVOID} cues,
    \item $c$ denotes optional execution constraints,
    \item $v$ denotes validation hooks or executable checks,
    \item $\eta$ denotes canonical metadata such as type, version, identifier, and hash.
\end{itemize}

This abstraction matches the operational schema used in our Gene-Bench setup, where a gene is serialized with fields such as \texttt{type}, \texttt{schema\_version}, \texttt{id}, \texttt{signals\_match}, \texttt{summary}, \texttt{strategy}, \texttt{constraints}, \texttt{validation}, and \texttt{asset\_id}. 
A representative example is a \texttt{<strategy-gene>} block containing domain keywords, a one-sentence summary, a short strategy list, and one or more \texttt{AVOID} items. 
In our experiments, such genes are typically around $230$ tokens and are designed to provide control-relevant experience under limited inference budget.

\subsection{Capsule Objects}
\label{app:gep_capsule}

A capsule records a validated task-solving path that can serve as a higher-level reusable asset. 
Unlike a gene, which abstracts a compact control strategy, a capsule preserves the successful realization of a strategy in context. 
For clarity, we formalize a capsule in this paper as
\begin{equation}
\kappa = (q, G_{\kappa}, T_{\kappa}, o_{\kappa}, V_{\kappa}, \ell_{\kappa}),
\qquad \kappa \in \mathcal{C},
\end{equation}
where
\begin{itemize}
    \item $q$ is a task signature or problem context,
    \item $G_{\kappa}\subseteq \mathcal{G}$ is the set of genes used or instantiated,
    \item $T_{\kappa}$ is the execution trace or action path,
    \item $o_{\kappa}$ is the observed outcome,
    \item $V_{\kappa}$ is the validation record or audit trail,
    \item $\ell_{\kappa}$ is the lineage pointer linking the capsule to prior events or parent assets.
\end{itemize}

This paper-side formalization is consistent with the public GEP description, in which capsules correspond to successful task execution paths and serve as validated assets above the gene layer. 
Conceptually, a capsule answers the question: \emph{How was a control strategy actually realized and validated in a concrete task episode?}

\subsection{Event Objects}
\label{app:gep_event}

Events are immutable records of protocol-relevant changes. 
They preserve provenance and make evolution auditable. 
We formalize an event as
\begin{equation}
e=(t,\rho,a_{\mathrm{src}},a_{\mathrm{dst}},\sigma,\iota,\Delta,\nu,\tau),
\qquad e\in\mathcal{E},
\end{equation}
where
\begin{itemize}
    \item $t$ is an event type (e.g., \texttt{repair}, \texttt{innovation}, \texttt{validation\_pass}, \texttt{validation\_fail}, \texttt{solidify}),
    \item $\rho$ is the parent run or episode identifier,
    \item $a_{\mathrm{src}}$ and $a_{\mathrm{dst}}$ denote source and target assets,
    \item $\sigma$ is the triggering signal,
    \item $\iota$ is the intended mutation or repair objective,
    \item $\Delta$ is the proposed change or diff,
    \item $\nu$ is the validation result,
    \item $\tau$ is the timestamp or ordering key.
\end{itemize}

This formalization instantiates the GEP view that events are immutable evolution logs recording each mutation or repair with associated context. 
Events are therefore not direct control representations; they are the protocol substrate for traceability, auditability, and lineage reconstruction.

\subsection{The GEP Loop}
\label{app:gep_loop}

The protocol updates experience objects through a six-stage loop:
\begin{equation}
(\mathcal{G},\mathcal{C},\mathcal{E}) 
\xrightarrow{\textsc{scan}} \sigma
\xrightarrow{\textsc{signal}} \hat{\sigma}
\xrightarrow{\textsc{intent}} \iota
\xrightarrow{\textsc{mutate}} a'
\xrightarrow{\textsc{validate}} \nu
\xrightarrow{\textsc{solidify}} (\mathcal{G}',\mathcal{C}',\mathcal{E}'),
\end{equation}
where $a'$ denotes a candidate asset, typically a new or revised gene (and optionally an associated capsule), and $\nu$ is the validation outcome.

\nbf{Scan.}
Runtime traces, tool logs, execution failures, or stagnation signals are monitored.

\nbf{Signal.}
Raw traces are converted into standardized protocol signals. 
This step turns unstructured execution evidence into machine-actionable mutation or repair triggers.

\nbf{Intent.}
The protocol determines the objective of evolution, such as repair, optimization, or extension.

\nbf{Mutate.}
A candidate asset is generated by rewriting code, prompt structure, strategy steps, \texttt{AVOID} cues, or associated metadata.

\nbf{Validate.}
The candidate is executed in a sandbox or otherwise checked against validation hooks. 
Only assets that satisfy the required criteria are eligible for persistence.

\nbf{Solidify.}
Validated capability is written back into the protocolized repository as a new or revised gene, and the corresponding capsule and event records are updated.

This loop is the protocol-level analogue of trial--validation--solidification. 
Its function is not merely to repair a single run, but to convert transient runtime adaptation into persistent reusable experience.

\subsection{Canonicalization and Protocol Invariants}
\label{app:gep_invariants}

For protocolization to be meaningful, experience objects must satisfy several invariants.

\nbf{Stable boundaries.}
A gene must have explicit fields and canonical serialization, so that it can be identified as an object rather than treated as free-form text.

\nbf{Control-oriented structure.}
A gene must encode compact control-relevant content, rather than documentation-heavy explanation. 
In our setting, this typically takes the form of matching signals, summary, strategy, and \texttt{AVOID} cues.

\nbf{Operability.}
Protocolized genes must support matching, replacement, revision, and composition.

\nbf{Validatability.}
A gene or capsule must expose executable or otherwise explicit validation interfaces.

\nbf{Lineage and auditability.}
Capsules and events must preserve enough provenance information to reconstruct how an experience object was produced, validated, and modified.

These invariants explain why GEP should be viewed as a protocol layer rather than a formatting convention. 
Canonicalization is what turns a useful prompt fragment into an analyzable, composable, and evolvable experience object.

\section{Detailed Experimental Setup}
\label{app:exp_setup}

Our goal is to document the retained experimental setup in a compact paper-facing form, focusing on evaluation, benchmark scope, inference configuration, and control representations.
All retained analyses share the same execution-based evaluation pipeline and prompt injection interface, so differences across conditions can be attributed primarily to the control representation rather than to changes in task collection, model family, or evaluation procedure.

\subsection{Evaluation Protocol and Metrics}
\label{app:eval_metric}

Each trial follows the same execution-based pipeline.
Given a task description and an optional control prompt, the model generates a Python program, the program is executed in a sandbox, and a scenario-specific test script evaluates a predefined set of checkpoints (sub-tasks).
Each execution is subject to a timeout of $120$ seconds.
\cref{tab:eval_protocol} summarizes the evaluation protocol and the reported metrics.

\begin{table}[t]
\centering
\caption{Evaluation protocol and reported metric.}
\label{tab:eval_protocol}
\begin{adjustbox}{max width=\columnwidth}
\begin{tabular}{ll}
\toprule
\textbf{Item} & \textbf{Definition} \\
\midrule
Unit of evaluation & One scenario--model--condition trial \\
Execution environment & Sandbox execution \\
Timeout & 120 seconds per trial \\
Primary metric & Pass rate (checkpoint-based) \\
Trial-level score & $n_{\mathrm{pass}} / n_{\mathrm{total}}$ \\
Secondary indicator & Complete pass ($n_{\mathrm{pass}} = n_{\mathrm{total}}$) \\
Relative metric & Improvement over baseline in pp \\
Aggregation & Average of trial-level pass rates across trials \\
\bottomrule
\end{tabular}
\end{adjustbox}
\end{table}

Formally, for trial $i$, let $p_i$ denote the number of checkpoints passed by the generated program and $n_i$ the total number of checkpoints for that scenario.
We define the trial-level pass rate as
\begin{equation}
r_i = \frac{p_i}{n_i}.
\end{equation}
For a condition with $N$ trials, the reported pass rate is
\begin{equation}
\mathrm{PassRate}
=
\frac{1}{N}\sum_{i=1}^{N} r_i
=
\frac{1}{N}\sum_{i=1}^{N}\frac{p_i}{n_i}.
\end{equation}
We report improvement relative to the no-guidance baseline in percentage points:
\begin{equation}
\Delta_{\mathrm{pp}}
=
\mathrm{PassRate}_{\mathrm{cond}}
-
\mathrm{PassRate}_{\mathrm{base}}.
\end{equation}

For completeness, we also record whether a trial achieves a complete pass:
\begin{equation}
\mathrm{Passed}_i = \mathbf{1}[p_i = n_i].
\end{equation}
This complete-pass indicator is used only as a secondary descriptive statistic when needed; throughout the paper, \textbf{pass rate} refers to the checkpoint-based metric defined above.
As summarized in \cref{tab:eval_protocol}, this design preserves partial success signal while preventing scenarios with larger test suites from dominating the aggregate score.

For brevity, later tables use the following notation:
\begin{itemize}[leftmargin=1.2em, itemsep=0.15em, topsep=0.2em, parsep=0pt, partopsep=0pt]
    \item \textbf{Pro} / \textbf{Flash}: Gemini 3.1 Pro / Gemini 3.1 Flash Lite.
    \item \textbf{Avg.}: arithmetic mean across the two models.
    \item \textbf{Avg.\ pass rate}: corresponding mean pass rate.
    \item \textbf{Avg.\ vs baseline} ($\Delta_{\mathrm{pp}}$): difference from the no-guidance baseline, in percentage points (pp).
\end{itemize}

\subsection{Benchmark Scope and Retained Analyses}
\label{app:benchmark_scope}

\cref{tab:scenario_scope} summarizes the benchmark scope used in this paper.
This study includes \TotalTrials{} retained trials over \TotalScenarios{} scientific scenarios.
Each scenario is paired with a scenario-specific execution script and evaluated at checkpoint granularity, allowing us to compare conditions on the same task collection while preserving partial success signal.
The benchmark spans a broad range of scientific and technical domains, including bioinformatics, neuroscience, chemistry, seismology, climate and atmospheric science, signal processing, network analysis, finance, robotics, and quantum computing.
Across these domains, tasks cover structured parsing and transformation, scientific analysis pipelines, simulation and fitting, algorithmic planning, and structured output generation.

\begin{table}[t]
\centering
\caption{Benchmark scope and scenario collection.}
\label{tab:scenario_scope}
\begin{adjustbox}{max width=\columnwidth}
\begin{tabular}{ll}
\toprule
\textbf{Item} & \textbf{Value} \\
\midrule
Total retained trials & \TotalTrials{} \\
Scenarios covered & \TotalScenarios{} \\
Scenario format & Self-contained task + scenario-specific test script \\
Evaluation granularity & Checkpoint-based within each scenario \\
Domain coverage & Broad scientific and technical domains \\
Representative task types & Parsing, analysis, simulation, planning, and structured output generation \\
\bottomrule
\end{tabular}
\end{adjustbox}
\end{table}

The retained analyses are organized into three groups.
\emph{Skill Probe} examines why documentation-oriented skills often fail to provide stable test-time control, including direct Skill-versus-Gene comparison, the effect of control length and token budget, and the contribution of individual skill components.
\emph{Gene Probe} asks whether genes function as a distinct experience representation rather than merely a shorter prompt, covering same-source comparison, robustness to content and structure perturbations, and limited complementarity with auxiliary context.
\emph{Evolution Probe} studies which representational properties make genes suitable as carriers of accumulated experience, including structured-versus-static presentation, failure-aware organization, ordering of failure and strategy information, and evolution-style narrative wrappers.

To make the benchmark concrete, we highlight three representative scenarios.
\texttt{S002\_spike\_behavior} is a computational neuroscience task with 12 checkpoints.
It requires the model to read neural spike times and behavioral velocity data from a MATLAB \texttt{.mat} file, filter successful trials, bin spikes into uniform time windows, align behavioral signals to bin centers by interpolation, and write a trial-structured HDF5 output.
Its checkpoints cover MATLAB cell-array parsing, successful-trial filtering, numerical correctness of spike binning, interpolation fidelity, HDF5 structural correctness, and quality-control logic for abnormal firing rates and NaN values.

\texttt{S026\_earthquake\_catalog} is a seismology task with 14 checkpoints.
It requires earthquake catalog analysis for aftershock identification, completeness magnitude estimation, Gutenberg--Richter $b$-value computation using the Aki maximum-likelihood formula, and structured CSV/JSON summaries.
Its checkpoints include datetime-aware parsing, Haversine distance computation, joint spatial-and-temporal aftershock identification, magnitude-frequency statistics, numerical correctness of the $b$-value estimate, and correctness of the summary outputs.

\texttt{S114\_obstacle\_avoidance} is a robotics and motion-planning task with 11 checkpoints.
It requires the model to construct a valid 2D trajectory using either RRT or a potential-field method, perform geometric collision checking, and report path-level metrics such as total length, waypoint count, obstacle clearance, and smoothness.
Its checkpoints test environment parsing, RRT initialization, nearest-node expansion, line-circle collision checking, path smoothing, potential-field force computation, and validity of the final path and metric report.

Taken together, these examples show that the benchmark varies substantially in domain semantics, algorithmic structure, and checkpoint granularity, which is why checkpoint-based evaluation is necessary for a faithful comparison of test-time control conditions.

\begin{table}[t]
\centering
\caption{Model and inference configuration.}
\label{tab:model_config}
\begin{adjustbox}{max width=\columnwidth}
\begin{tabular}{ll}
\toprule
\textbf{Configuration} & \textbf{Value} \\
\midrule
Strong model & Gemini 3.1 Pro Preview \\
Weak model & Gemini 3.1 Flash Lite Preview \\
API & Google AI Studio REST API \\
Decoding temperature & 0.05 \\
Max output tokens & 16384 \\
Control injection & \texttt{systemInstruction} \\
Task input field & \texttt{contents} \\
Execution timeout & 120 seconds \\
\bottomrule
\end{tabular}
\end{adjustbox}
\end{table}

\subsection{Model and Inference Configuration}
\label{app:model_config}

\cref{tab:model_config} summarizes the model and inference configuration used in the retained experiments.
All experiments use two fixed Gemini models: Gemini 3.1 Pro Preview and Gemini 3.1 Flash Lite Preview.
Generation is performed with low-temperature decoding ($T=0.05$), a maximum output budget of 16,384 tokens, and a shared control interface.
Control prompts are injected through the \texttt{systemInstruction} field, while the task description is supplied separately through the user-facing \texttt{contents} field.

\subsection{Control Representations and Injection Interface}
\label{app:prompt_templates}

We distinguish two main control representations in the retained analyses: \texttt{Gene} and \texttt{Skill}.
\texttt{Gene} is the compact strategy representation used throughout the main comparisons and is approximately 230 tokens long in the retained setting.
\texttt{Skill} denotes the full documentation-style skill package, including the main skill document together with auxiliary scripts, notes, and references, and is approximately 2,500 tokens long.
A no-guidance condition is also included as the baseline, in which no auxiliary control prompt is injected.

Across all conditions, the task specification is held fixed at the scenario level, while the control representation varies by experimental condition.
The task description is supplied through the user-facing \texttt{contents} field, and any auxiliary control context is injected separately through \texttt{systemInstruction}.
This separation ensures that comparisons are driven by differences in control representation rather than by rewriting the underlying task itself.

Gene conditions use a compact, strategy-oriented control representation.
In the retained experiments, the injected Gene content is serialized as a structured GEP-style strategy object and primarily contains concise strategic cues, such as domain-relevant signals, a short summary, and core strategy steps, rather than full documentation.
Skill conditions use a documentation-oriented control representation, in which the full skill package is provided as external guidance for the same underlying task.
Some retained analyses further place the Gene representation inside an evolution-style contextual wrapper, allowing us to study whether failure-aware organization and iterative experience framing change its effectiveness.

Representative task and control prompt examples are provided in~\cref{app:prompt_examples}.

\subsection{Representative Prompt Instances}
\label{app:prompt_examples}

This subsection provides representative prompt instances used in the retained analyses.
For a fixed scenario, the task prompt is held constant across conditions, while the auxiliary control prompt varies with the tested representation.

\nbf{Representative task prompt instance.}
Below is a representative task prompt instance based on the
\texttt{S002\_spike\_behavior} scenario:

\begin{verbatim}
Write a Python CLI script that reads neural spike times and behavioral
velocity data from a MATLAB .mat file and produces a trial-structured
HDF5 file.

The script must:
- filter to successful trials only
- bin spike times into uniform time windows
- resample behavioral velocity to match bin centers via interpolation
- write each trial as /trial_NNNN/spikes, /trial_NNNN/behavior,
  and /trial_NNNN/timestamps
- raise a quality-check flag for trials with firing rates above 200 Hz
  or NaN velocity values

The script must expose a command-line interface with:
--input, --output, --bin-size
\end{verbatim}

\nbf{Representative Gene control prompt.}
For Gene conditions, the task prompt is paired with a compact
strategy-oriented control prompt:

\begin{verbatim}
You are given the following strategic gene to guide your approach.
The gene describes a high-level strategy -- use it as directional guidance,
not as implementation instructions.

<strategy-gene>
Domain keywords: {keywords}
Summary: {summary}
Strategy:
  1. {step_1}
  2. {step_2}
  3. AVOID: {pitfall}
</strategy-gene>
\end{verbatim}

\nbf{Representative Skill control prompt.}
For the Skill condition, the same task prompt is paired with a
documentation-style full skill package:

\begin{verbatim}
You are given the following skill package to guide your work.
Follow its instructions carefully.

<SKILL.md + scripts + ...>
\end{verbatim}

\nbf{Representative evolution-style control prompt.}
Some retained analyses additionally wrap the Gene representation in an
evolution-style context that presents it as an iteratively refined
experience object:

\begin{verbatim}
This gene was evolved over multiple iterations.
The following evolution event captures what was learned:

<evolution-event intent="optimize" mutations_tried="4" outcome="success">
  <history>
    Attempt 1: [failure_1] -> FAILED
    Attempt 2: [failure_2] -> FAILED
    Attempt 3: [failure_3] -> FAILED
    Attempt 4: Applied learned strategy -> PASSED
  </history>
  <validated-gene>
    [gene_content]
  </validated-gene>
</evolution-event>
\end{verbatim}

These examples illustrate the separation between task specification and
control representation: the underlying scenario remains fixed, while the
auxiliary control context varies by experimental condition.

\section{Additional Experimental Details}

\subsection{Decomposing \texttt{Skill} into Sections}
\label{app:skill_sections}

\cref{tab:app_skill_components} reports the section-level decomposition of \texttt{Skill}. 
Starting from the full \texttt{Skill} document, we isolate individual sections to examine which parts carry usable test-time control signal. 
This analysis clarifies whether the utility of \texttt{Skill} is broadly distributed across the document or concentrated in a small subset of sections.

\begin{table}[h]
\centering
\small
\caption{Section-level decomposition of \texttt{Skill}. Avg. is the mean of Pro and Flash; $\Delta$ is relative to no guidance.}
\label{tab:app_skill_components}
\begin{tabular}{lcccc}
\toprule
Condition & Pro & Flash & Avg. & $\Delta$ \\
\midrule
\texttt{Skill-Overview} & 51.8\% & 40.8\% & 46.3\% & -4.7 \\
\texttt{Skill} & 50.7\% & 49.0\% & 49.9\% & -1.1 \\
No guidance & 60.1\% & 41.8\% & 51.0\% & 0.0 \\
\texttt{Skill-Pitfalls} & 55.3\% & 44.9\% & 50.1\% & +0.1 \\
\texttt{Skill-QuickRef} & 58.4\% & 44.6\% & 51.5\% & +0.5 \\
\texttt{Skill-ErrorHandling} & 56.2\% & 47.2\% & 51.7\% & +0.7 \\
\texttt{Skill-Workflow} & 52.6\% & 52.5\% & 52.5\% & +1.5 \\
\cellcolor{bestRow}\texttt{Gene} & \cellcolor{bestRow}59.9\% & \cellcolor{bestRow}48.2\% & \cellcolor{bestRow}54.0\% & \cellcolor{bestRow}+3.0 \\
\bottomrule
\end{tabular}
\end{table}

\subsection{Comparing \texttt{Gene} with Budget-Matched \texttt{Skill} Fragments}
\label{app:skill_budget_matched}

\cref{tab:app_skill_matched} reports the matched-budget comparison between \texttt{Gene} and shortened \texttt{Skill}-derived variants. 
For this comparison, we truncate the selected \texttt{Skill} fragments to approximately 230 tokens, so that they match the prompt budget of \texttt{Gene} as closely as possible. 
This analysis isolates whether \texttt{Gene}'s advantage can be explained primarily by brevity, or whether it also reflects a more effective control-oriented representation.

\begin{table}[h]
\centering
\small
\caption{Comparison between \texttt{Gene} and budget-matched \texttt{Skill} fragments. Avg. is the mean of Pro and Flash; $\Delta$ is relative to no guidance.}
\label{tab:app_skill_matched}
\begin{tabular}{lcccc}
\toprule
Condition & Pro & Flash & Avg. & $\Delta$ \\
\midrule
No guidance & 60.1\% & 41.8\% & 51.0\% & 0.0 \\
\texttt{Skill-Workflow-Short} & 54.6\% & 48.3\% & 51.5\% & +0.5 \\
\texttt{Skill-Pitfalls-Short} & 54.8\% & 49.2\% & 52.0\% & +1.0 \\
\cellcolor{bestRow}\texttt{Gene} & \cellcolor{bestRow}59.9\% & \cellcolor{bestRow}48.2\% & \cellcolor{bestRow}54.0\% & \cellcolor{bestRow}+3.0 \\
\bottomrule
\end{tabular}
\end{table}

\subsection{Additional Details on Content and Structural Perturbations}
\label{app:robustness_details}

This subsection provides additional details for~\cref{sec:robustness_content_structural}. 
Beyond the main contrast between content corruption and structural perturbation, the full results reveal a more nuanced pattern: some mutations remain effective even when they are no longer technically up-to-date, whereas others fail once they break semantic alignment with the target task.

\begin{table}[h]
\centering
\small
\caption{Additional details for content and structural perturbations of \texttt{Gene}. Avg. is the mean of Pro and Flash. Rows are ordered by Avg.}
\label{tab:app_gene_perturbations_full}
\begin{tabular}{lccc}
\toprule
Condition & Pro & Flash & Avg. \\
\midrule
Wrong algorithm & 49.6\% & 47.9\% & 48.8\% \\
Wrong domain & 52.0\% & 46.7\% & 49.4\% \\
Inverted priority & 54.8\% & 50.8\% & 52.8\% \\
Clean \texttt{Gene} & 59.9\% & 48.2\% & 54.0\% \\
Overconstrained & 57.2\% & 54.5\% & 55.9\% \\
\cellcolor{bestRow}Stale paradigm & \cellcolor{bestRow}59.9\% & \cellcolor{bestRow}53.4\% & \cellcolor{bestRow}56.6\% \\
\bottomrule
\end{tabular}
\end{table}

\nbf{An outdated solution can remain useful when it preserves the right problem framing.}
The most surprising result is \texttt{stale\_paradigm}, which reaches the highest Avg.\ at $56.6\%$, exceeding clean \texttt{Gene} at $54.0\%$. 
On Pro, it matches clean \texttt{Gene} in average pass rate ($59.9\%$), and on Flash it improves from $48.2\%$ to $53.4\%$. 
This result is not entirely implausible for difficult scientific scenarios. A reasonable prior expectation is that older methods may still retain experiential value, not because they remain technically optimal, but because they often encode stable ways of structuring the problem and narrowing the solution space. 
The current result is consistent with that interpretation: an outdated solution can remain effective when it still provides a task-relevant control framing.

\nbf{However, this robustness does not extend to semantically misaligned mutations.}
Both \texttt{wrong\_algorithm} and \texttt{wrong\_domain} substantially reduce performance, to $48.8\%$ and $49.4\%$, respectively. 
This sharp contrast helps refine the interpretation of~\cref{sec:robustness_content_structural}: \texttt{Gene} is not robust because incorrect content is harmless, but because some mutations preserve the task-relevant control structure even when their specific technique is suboptimal. 
Once the mutation breaks semantic alignment with the target task, the advantage disappears.

\nbf{Structural perturbation is tolerated more easily than content corruption.}
Both \texttt{inverted\_priority} and \texttt{overconstrained} remain competitive with clean \texttt{Gene}, reaching $52.8\%$ and $55.9\%$, respectively. 
This indicates that the effect of \texttt{Gene} is not tied to one fixed textual realization. 
The representation tolerates substantial variation in ordering and emphasis, provided that the encoded experience remains relevant to the task.

The effect of \texttt{Gene} is neither fragile with respect to wording nor indifferent to content quality. 
Rather, it depends on whether the mutated object preserves a task-appropriate control framing, even when the encoded method is somewhat outdated.

\section{Additional Details on Gene-Driven Test-Time Evolution}
\label{app:critpt_evolution_details}

We designed two evolution tasks on top of OpenClaw\footnote{\url{https://github.com/openclaw/openclaw}} and Evolver\footnote{\url{https://github.com/EvoMap/evolver}}, and let the resulting evolutionary agent run for approximately two days in each version.
The two tasks were intentionally different.
The first task targeted \emph{memory-grounded} self-improvement: the goal was to stabilize execution by repeatedly converting observed failures into reusable repair and protocol assets.
The second task targeted \emph{exploration-augmented} self-improvement: the goal was to expand the usable gene pool by combining prior run experience with externally sourced genes and topic-facing heuristics.
This setting matches Evolver's public framing of \emph{Exploration mode} as moving beyond single-loop self-repair toward proactive acquisition of new genes.

\subsection{Evolution Task Design for \texttt{Evolver (Gene) 2026-02-16}}
\label{app:first_run_design}

\nbf{Task objective.}
For \texttt{Evolver (Gene) 2026-02-16}, we designed an evolution task centered on \emph{stabilization through memory}.
The agent was not asked to broadly search for new capabilities.
Instead, it was asked to improve by looking inward: inspect prior logs, identify recurring execution failures and bottlenecks, and distill those observations into reusable genes.
The core hypothesis was that a useful evolution carrier should first be able to preserve and replay corrective experience, rather than merely produce one-off fixes.

\nbf{Evolution mechanism.}
Within this setting, Evolver acts as a structured evolution engine rather than a free-form prompting layer.
It selects or updates genes based on runtime signals, constrains candidate modifications through explicit strategy fields, and requires validation before successful experience is solidified.
As a result, the evolved agent does not simply ``remember more'' in a loose sense.
Instead, it gradually accumulates compact procedures that can be reinvoked in later runs.

\nbf{Why this version is memory-grounded.}
The first version is memory-grounded because the dominant improvement signal comes from prior failures and internal traces.
The agent evolves by repeatedly compressing observed breakdowns into reusable defensive procedures.
This makes the resulting gains structurally conservative but highly practical: the system becomes more stable not by adding broad new knowledge, but by turning past errors into future-time control.

\nbf{Cost profile.}
Completing the full benchmark with \texttt{Evolver (Gene) 2026-02-16} cost approximately $0.81\$$ when only observable input and output tokens are counted.
Reasoning-token costs could not be reliably recovered from our logs, so this number should be viewed as an input/output-only estimate.
Under this accounting, the run costs about $3.1\%$ of the reported $26\$$ benchmark cost (gemini-3-pro-preview), or roughly 32.1$\times$ less.

\subsection{Expanded Cases for \texttt{Evolver (Gene) 2026-02-16}}
\label{app:first_run_cases}

\nbf{Case A1: \texttt{gene\_gep\_repair\_from\_errors}.}
This gene is the clearest example of how the first version turns failure memory into reusable control.
It is triggered by explicit error-related signals such as \texttt{error}, \texttt{exception}, \texttt{failed}, and \texttt{unstable}.
Its strategy is procedural rather than rhetorical:
(1) extract structured signals from logs and instructions;
(2) select an existing matching Gene instead of improvising;
(3) estimate blast radius before editing;
(4) apply the smallest reversible patch;
(5) validate the result; and
(6) solidify the successful pattern back into the Gene/Capsule store.
This is important because the gene does not merely advise the agent to ``debug carefully.''
It stores a reusable repair loop:
\emph{structured diagnosis $\rightarrow$ constrained patching $\rightarrow$ validation $\rightarrow$ solidification}.
Once this loop is encoded as a gene, later runs can invoke it directly rather than rediscovering it from scratch.

\nbf{Case A2: \texttt{gene\_gep\_innovate\_from\_opportunity}.}
A complementary case is \texttt{gene\_gep\_innovate\_from\_opportunity}, which shows that the first version does not only repair failures, but also internalizes useful opportunity signals.
This gene is matched to signals such as \texttt{user\_feature\_request}, \texttt{user\_improvement\_suggestion}, \texttt{perf\_bottleneck}, \texttt{capability\_gap}, and \texttt{stable\_success\_plateau}.
Its strategy is again strongly procedural:
extract opportunity signals, search existing Genes and Capsules to avoid reinvention, design a minimal and testable implementation plan, estimate blast radius, validate the change, and finally solidify the new pattern as a Gene or Capsule if it proves useful.
Compared with the repair gene above, this gene is less reactive and more forward-looking.
However, it still remains memory-grounded in the sense that it derives new structure from accumulated runtime observations rather than from broad external exploration.
Together, Cases A1 and A2 show why gene iteration is already useful in the first version:
it preserves both defensive repair routines and internally discovered improvement routines as reusable evolution assets.

\subsection{Evolution Task Design for \texttt{Evolver (Gene) 2026-03-26}}
\label{app:second_run_design}

\nbf{Task objective.}
For \texttt{Evolver (Gene) 2026-03-26}, we designed the evolution task in the spirit of Evolver's \emph{Exploration mode}.
The goal was no longer only to consolidate existing repair knowledge, but to broaden the evolution substrate by acquiring and routing new task-relevant genes.
Concretely, this meant combining previously accumulated run experience with externally sourced genes, especially paper-derived and topic-facing assets, so that the agent could learn not only from past failures but also from possible future solution patterns.
The generated answers are publicly available at~\url{https://github.com/EvoMap/critpt-openclaw-reproducible-70}.

\nbf{Exploration-oriented mechanism.}
This version still keeps the base model fixed.
The change happens at the experience layer.
The agent is allowed to expand beyond the internal repair loop by selecting from a larger gene pool and routing different genes to different tasks.
In this sense, the second task is not merely a larger version of the first.
It reflects a different evolutionary regime: instead of asking how to make the current agent more stable within known behaviors, it asks how to enlarge the set of reusable procedures available to the agent.

\nbf{Full-run statistics.}
The resulting run covers $70$ tasks with $210$ gene slots and $36$ unique gene IDs.
The source distribution is highly asymmetric:
\texttt{arxiv}-derived genes account for $148$ selections, compared with $44$ from \texttt{run\_experience} and $18$ from \texttt{builtin\_topic\_prior}.
This shows that the second version is not simply replaying a small internal memory store.
It is operating with a broadened experience substrate in which external knowledge plays a major role.

\subsection{Representative High-Frequency Genes in \texttt{Evolver (Gene) 2026-03-26}}
\label{app:second_run_cases}

These cases illustrate that the evolved pool contains heterogeneous reusable experience objects, including run-derived procedural genes, built-in topic priors, and literature-derived abstractions.

\nbf{Case B1: \texttt{gene\_topic\_hamiltonian\_inverse\_design}.}
\begin{itemize}[leftmargin=1.2em, itemsep=0.15em, topsep=0.2em]
    \item \textbf{Tags:} \texttt{hamiltonian\_inverse\_design}, \texttt{many\_body\_spin\_chain}, \texttt{matrix\_numerical\_method}
    \item \textbf{Principles:}
    \begin{itemize}[leftmargin=1.2em, itemsep=0.1em, topsep=0.1em]
        \item Enumerate constraints first (e.g., commutation, normalization, operator ordering), then construct coefficients under those constraints, and finally verify the target terms.
        \item For many-body chain problems, decompose into local terms while preserving index consistency, in order to avoid bit-order and sign-convention errors.
        \item For numerical linear algebra tasks, prioritize stability and reproducibility, ideally with both symbolic and numerical verification.
    \end{itemize}
\end{itemize}
\textit{Interpretation.} This is a task-facing strategy gene that packages recurring solution procedures rather than merely recording prior difficulty.

\nbf{Case B2: \texttt{gene\_arxiv\_ebe24f551e12}.}
\begin{itemize}[leftmargin=1.2em, itemsep=0.15em, topsep=0.2em]
    \item \textbf{Tags:} \texttt{matrix\_numerical\_method}
    \item \textbf{Principles:}
    \begin{itemize}[leftmargin=1.2em, itemsep=0.1em, topsep=0.1em]
        \item Introduces a data-driven structural reduction of dynamic games by embedding offline-compiled best-response maps to remove nested optimization layers and derivative coupling.
        \item Validates the resulting method with large-scale Monte Carlo simulations, emphasizing both solution quality and computational efficiency.
        \item Provides mathematical guarantees that the reduced problem remains consistent with Nash-style equilibria under standard regularity and approximation conditions.
    \end{itemize}
\end{itemize}
\textit{Interpretation.} This case shows that evolution can import reusable structural abstractions from external literature, even when they originate outside the immediate task domain.

\nbf{Case B3: \texttt{gene\_topic\_seed\_many\_body\_spin\_chain}.}
\begin{itemize}[leftmargin=1.2em, itemsep=0.15em, topsep=0.2em]
    \item \textbf{Tags:} \texttt{many\_body\_spin\_chain}
    \item \textbf{Principles:}
    \begin{itemize}[leftmargin=1.2em, itemsep=0.1em, topsep=0.1em]
        \item For many-body chain problems, decompose into local terms while preserving index consistency, in order to avoid bit-order and sign-convention errors.
    \end{itemize}
\end{itemize}
\textit{Interpretation.} This case shows that built-in topic priors and run-derived experience can converge to nearly identical practical heuristics.

\nbf{Case B4: \texttt{gene\_arxiv\_d1aec630df87}.}
\begin{itemize}[leftmargin=1.2em, itemsep=0.15em, topsep=0.2em]
    \item \textbf{Tags:} \texttt{general\_quantum\_reasoning}
    \item \textbf{Principles:}
    \begin{itemize}[leftmargin=1.2em, itemsep=0.1em, topsep=0.1em]
        \item Combines structured search with experience-memory optimization to support dynamic reasoning.
        \item Introduces PE-EMP for real-time evolution of adaptive meta-prompts.
        \item Uses a memory-optimization agent to manage the experience library and improve reasoning efficiency.
    \end{itemize}
\end{itemize}
\textit{Interpretation.} Unlike the task-facing run-experience genes, this case represents a higher-level reasoning methodology gene.

Taken together, these cases show that the second evolution run stores reusable experience at multiple levels of abstraction, ranging from localized procedural heuristics to broader reasoning frameworks.

\end{document}